# Coarse-graining algorithms for the Eulerian-Lagrangian simulation of particle-laden flows

H. Eshraghi, E. Amani[*], M. Saffar-Avval

*Department of Mechanical Engineering, Amirkabir University of Technology (Tehran Polytechnic), Iran*

**Abstract**

In the present article, novel Coarse-Graining (CG) algorithms for the Eulerian-Lagrangian (EL) simulation of particle-laden flows are proposed. These include different variants of Reproducing Kernel Particle Methods (RKPM) and an extended Diffusion Two-Step Method (DTSM) for highly polydisperse flows. Owing to the dynamic nature of the kernel function in RKPMs, CG algorithms with high-order consistency properties are constructed and the extra physics of the fluid-particle interaction torque effect on the two-way coupling force distributed to the fluid is taken into consideration. To increase the robustness of RKPMs, they are hybridized with other simple CG algorithms in such a way that each model is activated in the range of its validity. The performance of the new CG algorithms is carefully assessed in comparison to other widely-used algorithms by devising four benchmarks and several grid-independence tests. Based on the analyses, the first-order RKPM demonstrates the best performance in terms of 8 attributes, including conservativity, grid-independency, smoothness, etc., among all algorithms and is recommended when the accuracy is of prime importance. However, due to its higher computational cost compared to the extended DTSM, the latter model would be an affordable alternative for large-scale discrete element method problems when the computational cost is critical.

---

[*] Corresponding author. Address: Mechanical Engineering Dept., Amirkabir University of Technology (Tehran Polytechnic), 424 Hafez Avenue, Tehran, P.O.Box: 15875-4413, Iran. Tel: +98 21 64543404. Emails: eamani@aut.ac.ir (E. Amani), h.eshraghi@aut.ac.ir (H. Eshraghi), mavval@aut.ac.ir (M. Saffar-Avval).

# 1. Introduction

In Eulerian-Lagrangian (EL) particle-laden flow simulations, fluid properties at a particle location are required which are computed using interpolation algorithms. The reverse process, in which the particle properties are distributed over the Eulerian grid, is called Coarse-Graining (CG), backward interpolation, or averaging. CG is a key element of two-way coupled EL simulations or particle statistics calculation in the post-processing stage. In 1977, Crowe et al. [1] introduced the Particle Centroid Model (PCM), which is the simplest CG algorithm. This model is sometimes referred to as the Particle-In-Cell (PIC) model. The PCM model assigns the properties of a particle to the host cell in which the particle center resides. The PCM method suffers from several serious issues, especially when the particle diameter to cell size ratio, $d_p/\Delta$, is of the order of unity or larger. The unphysically large coarse-grained field values (boundness problem) and/or gradients and highly grid-dependent particle statistics are among the main drawbacks of PCM. Many CG models have been proposed to overcome the shortcomings of PCM, which can be classified into six general groups: Node-based, divided-particle-volume, Approximate divided-particle-volume, kernel-based, two-grid, and diffusion two-step models.

The Node-Based Methods (NBM) refer to a particular class of CG models in which the nodes and vertices of the host cell and its adjacent cells are employed to distribute particle properties over the Eulerian grid. The well-known trilinear scheme is one of the NMBs which is applicable to structured grids. In this model, each particle property is divided among 4 vertices in 2D [2] or 8 vertices in 3D [3] of the host cell. Then, these shares are dispersed throughout all the cell neighbors that have joint vertices with the host. An extension of this method to unstructured grids, termed the Grid Distribution Method (GDM), is implemented in the present work using the Barycentric coordinates. The present GDM is, in fact, the inverse of the procedure developed by Ebert and Smith [4] to map data between two unstructured grids. Although this model significantly reduces the fluctuations and noises of coarse-grained fields compared to PCM, it fails to produce grid-independent results when $d_p/\Delta \geq 1$.

For large particles ($d_p/\Delta \geq 1$), PCM or NBM may result in severe inaccuracies in estimating particle volume fraction or other coarse-grained fields. To overcome this issue, a set of models were presented in which the particle statistics are calculated using geometrical considerations. These models are called Divided-Particle-Volume Methods (DPVM), exact, or analytical methods. For 2D problems, a DPVM for cylindrical particles was proposed by Li [5]. For 3D cases, Peng et al. [6] formulated different scenarios of spherical particles inside a structured grid. Wu et al. [7] extended the DPVM to 3D unstructured grids with tetrahedral cells. They expressed the contribution of each particle to cells based on particle and cell vertices positions and in terms of complicated integral equations. The high computational cost of computing the integral equations adds a further complication to the already challenging task of identifying particle-cell arrangement scenarios. In addition, the model is applicable if $d_p/\Delta$ is not much larger than unity, more specifically if at least one edge of the cell is larger than the particle diameter.

To alleviate these problems, Approximate DPVM (ADPVM), also called the porosity mapping method, is an alternative. In 3D ADPVM, a particle is replaced with a cube of the same volume [8]. This assumption renders the cell shares of each particle much more easily computable. Link [9] reported that the volume fraction calculation can destabilize the two-way coupled simulations if a cell is almost entirely occupied by particles (solid volume fractions greater than about 0.95). Tomiyama et al. [10] and Link et al. [11] proposed a modification to ADPVM to overcome this issue. They suggested that the cube volume can be greater than the particle volume or alternatively the cube edge length is assumed as $d_{\text{cube}} = d_p a$, where the calibration coefficient $a$ must never be less than $(\pi/6)^{1/3}$. Link [9] suggested a value of 5 for $a$. To preserve the conservativity, the particle properties are distributed by reduced weights, named porosity. Increasing the cube volume may create an overlap between the cube and boundary walls. To handle these situations, a wall treatment method was proposed by Link [9], termed the folding technique, in which the portion of the cube volume which is located outside the domain beyond a wall boundary is mirrored back into the domain.

The ADPVM can be regarded as the precursor of Kernel-Based Methods (KBM) [12]. In KBM, the particle properties are distributed over a (usually) spherical region, called the Influence Domain (ID), in contrast to a cube in ADPVM. In addition, in KBM, the cells closer to a particle receive greater shares of the particle properties, while the particle properties in ADPVM are distributed uniformly. For this purpose, a kernel function, frequently referred to as the template window function, is used to account for the influence of particle-to-cell distance. Kitagawa et al. [13] pointed out that the kernel functions should meet the following five requirements: normalization, smoothness, compactness, descending, and maxima-at-zero. Different kernels have been used in EL simulations. Sun and Xiao [14] utilized an exponential function, Glasser and Goldhirsch [15] and Zhoo and Yu [16] used the Johnston function, and Kitagawa et al. [13] adopted a sinusoidal template window function. Evrard et.al [17] recommended the polynomial Wendland kernel due to its higher control over smoothness and relatively low computational cost. A clipped quadratic polynomial function was also used by Deen et al. [18] and Hu et al. [19]. To compute each cell weight more accurately, the kernel function should be integrated over the overlap region of a particle ID and each neighboring cell in the particle ID. This is a tedious task in 3D unstructured grids of arbitrary cell shapes. To extend KBM to unstructured grids, Hu et al. [19] suggested that the kernel function can be computed at cell centers and be assumed constant over each cell. This assumption has been used in many unstructured solvers, though it can degrade the estimation of cell weights. Sun and Xiao [14] used the Gaussian Kernel Method (GKM) and recommended an ID diameter of 12-18 times larger than the particle diameter, based on the wake behind a sphere predicted by a resolved approach. Like ADPVM, KBM requires special boundary treatment. Zhu and Yu [16] proposed a boundary treatment method, called the method of images or virtual particles, when the particle ID intersects single or multiple boundary walls. The virtual particles are considered as the reflection of real particles on the other side of the wall. However, these kinds of wall treatments are complex and involve inefficient recursive coding. Interested readers can consult reference [20] for additional information on wall treatment methods in CG algorithms.

Deb and Tafti [21] conceptualized an alternative approach, called Two-Grid Method (TGM), in which fluid and particles are handled on separate computational grids. The continuous Eulerian phase is solved with a fine grid while a coarse grid, which is generated by the agglomeration of the Eulerian cells, is applied for the Lagrangian phase. To guarantee that there is a sufficient number of particles in a Lagrangian cell, Muller et al. [22] stated that the Lagrangian cells are to be three times larger than the particle diameter. The particle properties are estimated on the Lagrangian mesh and evenly dispersed into the Eulerian sub-cells. Although this is an effectual approach in reducing the sharp unphysical gradients and noises, the parallel efficiency is a matter of concern due to the need for the agglomeration process across the processor boundaries.

A more recent group of models, called the Diffusion Two-Step Method (DTSM), which is based on the solution of diffusion equation on the Eulerian grid to disperse the particle properties was proposed by Capecelatro and Desjardins [23]. For the first step, they used the GKM with an ID proportional to the local Eulerian grid size. On a uniform orthogonal mesh, this corresponds to transferring particle effect to the 27 nearest cells. In the second stage, the coarse-grained field is further dispersed over the Eulerian grid by solving a diffusion equation initialized by the field obtained in the first step. The diffusion coefficient was determined based on the particle diameter as well as the grid length scale to remove the dependence of the result on the grid resolution.

Though this idea was a breakthrough to reduce the cost of CG and received a lot of attention from researchers, it had a number of difficulties. First, the first stage, i.e., the GKM, was a computationally expensive stage that hindered the efficiency of the model, especially on unstructured grids. Second, the application of the model to polydisperse flows was not straightforward. To solve the first issue, Sun and Xiao [24] adopted the PCM for the first stage and proposed an analytical diffusion coefficient in such a way that DTSM becomes an estimation of the GKM with a much lower computational cost. This modified DTSM is easier to implement and needs no extra wall treatments in contrast to GKM. However, as it is indicated in the present work, the modified DTSM needs a higher grid resolution to achieve grid-independent results. In addition, the extension of the model to polydisperse flows is still an open question.

In the present research, DTSM is extended to highly polydisperse flows. More importantly, in an effort to develop improved CG algorithms, a new category of models, called Dynamic KBM (DKBM), is proposed for the first time for CG in EL simulations (section 3), based on the so-called Reproducing Kernel Particle Methods (RKPM) originally developed by Liu et al. [25] for interphase force distribution in Immersed Boundary Methods (IBMs). Here, the word "dynamic" implies the variation or correction of the kernel function locally based on the grid and/or physics of the problem. The dynamic variation of the kernel allows the incorporation of additional desirable kernel numerical characteristics and even extra physics into EL simulations. Here, three variants of RKPM are presented, including the first- and second-order-consistent RKPMs and an RKPM variant for including the effect of drag torque in the two-way coupling force on the continuous phase, which has been always omitted in the previous EL simulations, to the best of the authors' knowledge. The performance of the new CG algorithms is carefully compared with other widely-used models, including PCM, GDM, GKM, and DTSM, via four different benchmarks devised in the present work (section 5). Finally, the conclusions and recommendations for the use of CG algorithms are provided (section 6).

## 2. Governing equations

In this section, the equations used in this study to govern the physics of a two-way coupled particle-laden flow are introduced. The particle and continuous-fluid phases are expressed by the Lagrangian and Eulerian approaches, respectively.

*2.1. Lagrangian equations*

Conforming to the Newton's second law, the motion of particles is given by:

$$\frac{d\boldsymbol{x}_p}{dt} = \boldsymbol{u}_p \tag{1}$$

$$m_p \frac{d\boldsymbol{u}_p}{dt} = \boldsymbol{F}_p + m_p \boldsymbol{g} \tag{2}$$

where $\boldsymbol{x}_p$ is the particle position, $\boldsymbol{u}_p$ the particle velocity, $m_p$ the particle mass, $t$ the time, $\boldsymbol{g}$ the gravitational acceleration, and $\boldsymbol{F}_p$ the particle-fluid interfacial force exerted on the particle. Here, the bold symbols indicate tensorial quantities. For the sake of simplicity, only the gravity and drag interfacial forces are considered in this paper, for the latter the Gidaspow (Ergun-Wen-Yu) correlation [26] is used, where Ergun's relation [27] is adopted when the void fraction is below 0.8 and the Wen-Yu relation [28] is used otherwise:

$$\boldsymbol{F}_p = \frac{V_p \beta}{1 - \theta_c}(\boldsymbol{u}_c - \boldsymbol{u}_p)$$

$$\beta = \begin{cases} 150 \frac{(1-\theta_c)^2}{\theta_c} \frac{\mu_c}{d_p^2} + 1.75(1-\theta_c)\frac{\rho_c}{d_p}|\boldsymbol{u}_c - \boldsymbol{u}_p| & ; \theta_c \leq 0.8 \\ \frac{3}{4}\frac{(1-\theta_c)}{d_p}\rho_c|\boldsymbol{u}_c - \boldsymbol{u}_p|C_D \theta_c^{-1.65} & ; \theta_c > 0.8 \end{cases} \tag{3}$$

where $d_p$ and $V_p$ are the particle diameter and volume, respectively. $\theta$ is the volume fraction, $\boldsymbol{u}$ the velocity, $\mu$ the viscosity, and $\rho$ the density. The particle and continuous-phase properties are indicated by subscripts $p$ and $c$, respectively. The continuous-phase properties values in particle equations are interpolated values at the particle center. The drag coefficient, $C_D$, is calculated here by the Schiller-Naumann relation [29]:

$$C_D = \begin{cases} \dfrac{24}{Re_p}[1 + 0.15 Re_p^{0.687}] & ; Re_p \leq 1000 \\ 0.44; & ; \text{otherwise} \end{cases} \quad (4)$$

where $Re_p$ is the particle Reynolds number given, for the Gidaspow correlation, as:

$$Re_p = \frac{\theta_c |\boldsymbol{u}_c - \boldsymbol{u}_p| d_p}{v_c} \quad (5)$$

and $v$ is the kinematic viscosity.

*2.2. Eulerian equations*

The governing equations for the continuous-phase flow are based on the Eulerian averaged approach. Under the assumption of no interfacial mass transfer, the equations can be derived as follows [21, 30-32]:

$$\frac{\partial}{\partial t}(\rho_c \theta_c) + \boldsymbol{\nabla}.(\rho_c \theta_c \boldsymbol{u}_c) = 0 \quad (6)$$

$$\frac{\partial}{\partial t}(\rho_c \theta_c \boldsymbol{u}_c) + \boldsymbol{\nabla}.(\rho_c \theta_c \boldsymbol{u}_c \boldsymbol{u}_c) = -\boldsymbol{\nabla} p + \boldsymbol{\nabla}.\theta_c \boldsymbol{\tau}_c + \rho_c \theta_c \boldsymbol{g} - \boldsymbol{f}_{FP} \quad (7)$$

where $p$ is the fluid pressure, $\boldsymbol{\tau}_c$ the viscous stress which is defined by $\boldsymbol{\tau}_c = v_c(\boldsymbol{\nabla}\boldsymbol{u}_c + (\boldsymbol{\nabla}\boldsymbol{u}_c)^T)$, $\boldsymbol{f}_{FP}$ the fluid-particle interaction force per unit volume exerted by the fluid on particles, and $\theta_c = 1 - \theta_s$ is obtained knowing the particle phase volume fraction, $\theta_s$. To close the equations, a CG algorithm should be chosen to compute $\theta_s$ and $\boldsymbol{f}_{FP}$, given each particle volume, $V_p$, and interfacial force, $\boldsymbol{F}_p$, introduced in section 2.1.

## 3. Coarse-graining algorithms

### 3.1. General considerations

Particle properties are transformed into Eulerian cells via CG models. An ideal CG algorithm possesses the following characteristics:

1- *Conservativity*: It insures the conservative distribution of properties from a particle to the Eulerian grid. In KBM CG algorithms, it is equivalent to the unity of the volume integral of the kernel function in the discretized form.

2- *Grid independency*: The sensitivity of the CG algorithm results to the choice of Eulerian grid should be minimal.

3- *Boundness*: Being bounded by a specified threshold is desirable for some properties like $\theta_s$, e.g., $\theta_s < \theta_s^* \sim 0.9$, for the stability of the EL simulation.

4- *Bandwidth adjustability*: It is desirable to distribute a particle property over an adjustable distance of the particle, which is usually determined based on the particle diameter.

5- *High-order consistency*: The first and/or higher moments of the distributed property may need to be consistent with the original one.

6- *Robustness*: The numerical instabilities of the solution algorithm should be minimized.

7- *Smoothness*: The physical and numerical considerations usually demand a smooth distribution of particle properties away from the particle.

8- *Boundary treatment*: Simple boundary treatment is desirable. Some algorithms require complex amendments near the domain boundaries.

9- *Computational cost*: A low computational overhead is a plus. The parallel efficiency is another computational concern.

In this section, some well-known CG algorithms along with the newly proposed ones are introduced, and the degree to which they satisfy the aforementioned attributes are discussed. The formulations are written for the transformation of a particle property, $\Phi_p$, into the corresponding Eulerian field property (per unit volume), $\phi$. In case of the particle volume fraction, $\Phi_p = V_p$ and $\phi = \theta_s$, and for the interfacial force, $\Phi_p = \boldsymbol{F}_p$ and $\phi = \boldsymbol{f}_{FP}$. The CG operation used in the present work can be written in the following general form:

$$\phi(x_j) = \sum_{p=1}^{N_{pj}} \phi_{pj} \qquad (8)$$

where $x_j$ is the location of the center of $j^{th}$ Eulerian cell, $N_{pj}$ is the number of particles in the region of influence of $x_j$, and $\phi_{pj}$ is the contribution of $p^{th}$ particle properties to the coarse-grained properties at $j^{th}$ cell which is determined based on one of the CG models introduced in the following sections.

### 3.2. Particle Centroid Method (PCM)

The PCM is the most straightforward and extensively used CG approach, where the properties transformation is performed by Eq. (8) using:

$$\phi_{pj} = \Phi_p/V_j \qquad (9)$$

where $V_j$ is the volume of cell $j$ and the summation in Eq. (8) is taken over all particles whose centers reside in cell $j$. The most significant advantage of the PCM is its simple implementation compared to the other models. The particle effect is only assigned to the host cell containing the particle center. The model accuracy is highly dependent on the ratio of the particle diameter to cell length scale, $d_p/\Delta$. For a small $d_p/\Delta$ (called here Scenario I), the PCM delivers satisfactory results. However, as shown in figure 1, when $d_p/\Delta \approx 1$ and the particle center is located near the boundaries of the cell (Scenario II), all particle contribution is still assigned to the host cell. This may impose a relative error of up to 50% in calculating $\theta_s$ at the host cell [6]. In Scenario III, the diameters of the particles are comparable to the cell length scale, and many particles accumulate in one cell in such a way that $\theta_s$ of the host cell unphysically exceeds unity. Similarly, for large $d_p/\Delta$ (Scenario IV), the host cell $\theta_s$ is overestimated, while the ones of the surrounding cells take the value of zero. This severe volume fraction change can lead to the solution instability. The procedure of implementing PCM is demonstrated in Algorithm 1 of Appendix A.

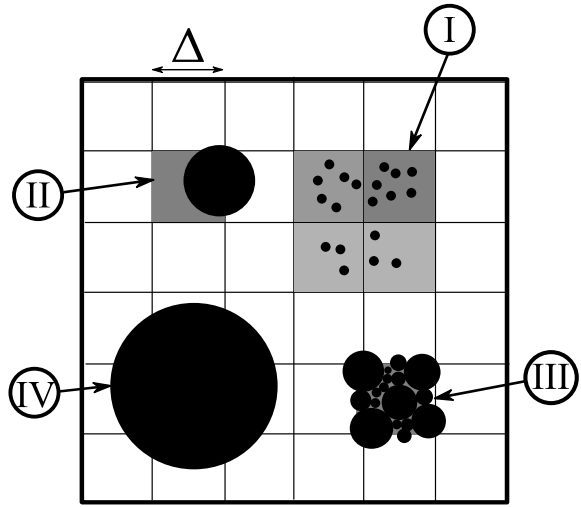

**Figure 1** Different scenarios based the particle sizes and locations in the Eulerian grid. Scenario I: $d_p/\Delta < 1$, Scenario II: $d_p/\Delta \approx 1$ and the particle center falls near the cell boundaries. Scenario III: $d_p/\Delta \approx 1$ and many particles accumulate in the same cell, and Scenario IV: $d_p/\Delta > 1$.

*3.3. Grid Distribution Method (GDM)*

The GDM is an easy-to-implement, low-cost computational model introduced here to alleviate some of the drawbacks of PCM. This model can be regarded as the extension of the trilinear model to 3D unstructured grids. The particle effect in the GDM is distributed in one direction, therefore, only the neighbor cells of the host which are closer to the particle receive the particle contributions. The model consists of two steps which are summarized as follows.

The first step involves the decomposition of the host cell into irregular tetrahedrons. Each tetrahedron shares three vertices with the host cell, and its fourth vertex is located at the center of the host cell. The particle contributions are received by the four vertices of the tetrahedron where the particle is located. This stage is, in fact, the inverse of the procedure proposed by Ebert and Smith [4] to map data between two unstructured grids. A particle property, $\Phi_p$, is divided into four shares at the corners of the host tetrahedron, i.e., at vertices ($\Phi_p^n; n = 1,2,3$) and the host cell center ($\Phi_p^c$), based on the weight coefficients ($\omega_1, \omega_2, \omega_3, \omega_c$):

$$\Phi_p^n = \omega_n \Phi_p, \quad \Phi_p^c = \omega_c \Phi_p \tag{10}$$

This step needs the transformation of the particle location from the global Cartesian coordinates to the host tetrahedron Barycentric coordinates [33] to determine the weight coefficients, $\omega_1$, $\omega_2$, $\omega_3$, and $\omega_c$ which are equal to the particle location coordinates in the Barycentric coordinates system of the host tetrahedron. Eq. (10) is indicated in line 5 of Algorithm 2 (Appendix A).

In the next step, the shares on the three cell vertices from the previous step are distributed to the surrounding cells that satisfy the following criteria: a) Sharing at least one vertex from vertices $n = 1,2,3$ with the host cell, and b) is a neighbor to the host cell (have a joint face). The property distribution is performed based on the inverse distance between the vertex and the neighboring cell centers. Finally, the contribution of particle $p$ to cell $j$, $\phi_{pj}$, in Eq. (8) is computed by:

$$\phi_{pj} = \frac{1}{V_j}\left(\Phi_p^c(j) + \sum_{n=1}^{N_{pn}} a_{p,j}^n \Phi_p^n(j)\right) \tag{11}$$

$$a_{p,j}^n = \frac{1}{|C_j - x_n|} \Big/ \sum_{j=1}^{N_{pnj}} \frac{1}{|C_j - x_n|} \tag{12}$$

where $C_j$ is the center of cell $j$, $N_{pn}$ ($\leq 3$) is the number of vertices of cell $j$ which receives a contribution from $\Phi_p$, and $N_{pnj}$ is the number of neighbor cells sharing the vertex $n$ with the host cell. $\Phi_p^c(j)$ is non-zero only for the host cell. Line 11 of Algorithm 2 (Appendix A) demonstrates

Eq. (11) and line 10 indicates Eq. (12). The first step of algorithm is described in lines 5 and 6, and lines 7 to 14 are the second step. For the sake of clarity, the representation of the algorithm steps is provided in figure 2.

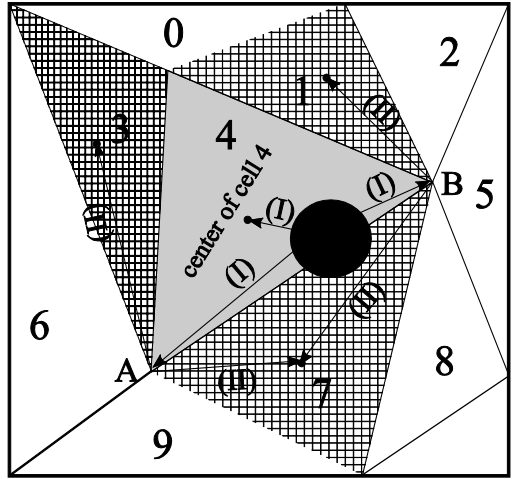

**Figure 2** The GDM algorrithm: The first (I) and second (II) steps of properties distriburion from a particle (black circle) to the Eulerian cells in a 2D unstructured grid.

*3.4. Gaussian Kernel Method (GKM)*

In the GKM, the contribution of particle $p$ to cell $j$, $\phi_{pj}$, in Eq. (8) is computed by:

$$\phi_{pj} = a_p \Phi_p \, g_p(x_j - x_p) \tag{13}$$

where the kernel of particle $p$ is denoted by $g_p(x)$, and $a_p$ is the renormalization coefficient of particle $p$ to ensure conservativity and is calculated by:

$$a_p = 1 / \left( \sum_{j=1}^{N_j} g_p(x_j - x_p) V_j \right) \tag{14}$$

where $N_j$ is the number of cells in the ID of particle $p$. Here, a Gaussian kernel is employed as:

$$g_p(x) = \frac{1}{(b^2 \pi)^{N_{\dim}/2}} \exp\left(-\frac{x \cdot x}{b^2}\right) \tag{15}$$

where $b$ is the kernel bandwidth which can be adjusted based on the particle size and $N_{\dim}$ is the number of problem dimensions which equals 3 here. Although the choice of ID size has been widely discussed in many papers, still many questions remain unanswered [16]. Evrard et al. [17] proved that the Wendland template function [34] provides a closely resembling substitute for the Gaussian function and choosing a $\delta_p$ (the ID diameter of particle $p$) greater than $2.41 d_p$ guarantees the model boundness. Link et al. [9] suggested $\delta_p/d_p = 5$, while $\delta_p/d_p = 3$ was chosen by Bokkers et al. [35] and $\delta_p/d_p = 8$ by Zhu and Yu [16]. A more physically-grounded approach was followed by Sun and Xiao [14], correlating the ID size to the wake behind particles. Based on this analysis, they suggested a value of 24 to 36 for $\delta_p/d_p$. Since large IDs considerably increase the cost of the search algorithm, in the present work, $\delta_p/d_p = 12$ is chosen as a compromise between accuracy and computational cost. For the Gaussian template function, $\delta_p = 6b$ [14], which necessitates the use of $b = 2d_p$ in Eq. (15) for the present simulations.

A detailed description of GKM is given in Algorithm 3 of Appendix A. The GKM algorithm is divided into two parts: The first part (lines 5 to 18) is responsible for the calculation of the correction factor $a_p$. If the near-wall treatment is enabled, the locations of virtual particles are obtained in line 8 by calling Algorithm 7 (discussed in section 3.8). In the second part, the cell contribution is collected using the pre-calculated correction factor in lines 19 to 30.

For the GKM and every other model using an ID, finding cells that are inside the region of influence of a particle is another challenge. An efficient approach is adopted here by finding and storing all cells which are in the region of influence of all possible locations within each cell, at the beginning of the simulation. This approach is described in Algorithm 8 of Appendix A. In this procedure, at beginning of the solution, the maximum ID size is calculated based on the largest particle diameter, i.e., $\delta_{\max} = R_b d_{p,\max}$ where $R_b$ is the ratio $\delta_p/d_p$ (which is chosen to be 12 in the current study by default). The local ID diameter for all locations within each Eulerian cell, $\delta_c$, which is called here the "local base ID" diameter, is then computed by adding the local grid length scale $\Delta_c$ to $\delta_{\max}$ in line 7 of the algorithm. This guarantees the local ID at each cell $c$ includes all neighboring cells which are inside the "local base ID". Then, for each cell $c$, the cells within the spheres of diameter $\delta_c$ centered at all vertices of cell $c$ are found and collected in a list (lines 8 to 12). Since this list contains duplicate cells, in line 13, the duplicate cells are found and removed. Note that a simpler approach can be conceived in which instead of looping over all vertices of cell $c$ and collecting the cells in their sphere domain, only the cells in a single sphere domain centered at the center of cell $c$ are sought. However, we found that this approach may miss some cells in the ID of a particle located in the corners of cell $c$, in the case of poor-quality grids with high aspect ratios. Since this part of the search, i.e., Algorithm 8, is carried out only once at the begging of the simulation, we adopted the more complicated former approach. The ID of each particle during the simulation is then confined to the "local base ID" of the host cell where the center of the particle resides. As a result, during the simulation, the search to find the cells in the ID of a particle is carried out within the "local base ID". This approach effectively reduces the computational cost of finding the cells in the ID of a particle during the simulation, as shown in section 5.4.

*3.5. Reproducing Kernel Particle Method (RKPM)*

RKPM is derived based on the technique applied for the force distribution in the IBM in the literature [25, 36]. In these works, the particle-fluid interface is covered with a number of points named "markers". These markers are responsible for transferring the interfacial force to the surrounding fluid cells. The moments of the distributed force by any marker to cells around the marker center is zero. In the present work, the RKPM for the dispersed-phase particles is constructed using the same procedure. Here, the Lagrangian particles are assumed to be markers, and a particle property is distributed by Eq. (8) where $\phi_{pj}$ is given by:

$$\phi_{pj} = \Phi_p \, \tilde{g}_p(x_j - x_p) \tag{16}$$

where $\tilde{g}_p(x)$ is the RKPM kernel for particle $p$. For a uniform Cartesian grid, $\tilde{g}_p(x) = g_p(x)$, where $g_p(x)$ is reproduced by the product of the 1D kernel functions in the three coordinates directions as:

$$g_p(x) = g_p(x)g_p(y)g_p(z) \tag{17}$$

and function $g_p(x)$ is usually chosen as the cosine form suggested by Peskin [37]:

$$g_p(x) = \begin{cases} \frac{1}{4b}\left(1 + \cos\left(\frac{\pi|x|}{2b}\right)\right) & ; |x|/b < 2 \\ 0 & ; \text{otherwise} \end{cases} \tag{18}$$

The domain of influence of Eq. (18) is $-2b < x < 2b$, thus, $\delta_p = 4b$, and given $\delta_p$, the bandwidth $b$ can be computed by $b = \delta_p/4$. In the IBM, $g_p(x)$ is considered to be an approximation to the 1D Dirac delta function; therefore, its bandwidth $b$ is chosen to be proportional to the grid cell size, $\Delta$. For the IBM on orthogonal and quasi-orthogonal grids in which a rectangular ID has been suggested by Pinelli et al. [38], the ID around the marker is selected in such a manner that in every direction, at least three nodes of the Eulerian grid are enclosed within the ID, while for non-orthogonal grids, an ID with a diameter of four times the host cell length scale was recommended by Sigüenza et al. [36]. For CG in EL simulations, a kernel with an adjustable bandwidth based on the particle size is desired. To be consistent with GKM, $\delta_p/d_p = 12$ is adopted for the present RKPMs. Note that, instead of Eq. (17), an alternative extension of the 1D weight function to higher dimensions can be obtained by multiplying Eq. (18) by $1/2^{N_{dim}-1}$ [36].

Sigüenzaa et al. [39] found that $\tilde{g}_p(x) = g_p(x)$ is not valid for general non-uniform or unstructured grids. They multiplied $g_p(x)$ by a corrector function, $C_p(x)$, to minimize the grid non-uniformity or distortion effect as:

$$\tilde{g}_p(x) = C_p(x)g_p(x) \tag{19}$$

For a general grid, the corrector function depends on the location of the particle within the Eulerian grid and should be calculated dynamically at every update of the grid or particle location. Here, two variants of RKPM are adopted. In the 1st-order RKPM (RKPM1), the corrector function is assumed as:

$$C_p(x) = \beta_0 + \beta_1 x + \beta_2 y + \beta_3 z \tag{20}$$

where the four coefficients $\beta_0$ to $\beta_3$ are computed so as to satisfy four constraints. These constraints are the conservativity criterion (property 1 in section 3.1) and the consistency of the first-order moments, or consistent torque in the case of $\Phi_p = F_p$, (property 5 in section 3.1). For this purpose, a system of linear equations is solved for each particle to find the four coefficients $\beta_0$ to $\beta_3$:

$$\begin{pmatrix} m_{0,0,0} & m_{1,0,0} & m_{0,1,0} & m_{0,0,1} \\ m_{1,0,0} & m_{2,0,0} & m_{1,1,0} & m_{1,0,1} \\ m_{0,1,0} & m_{1,1,0} & m_{0,2,0} & m_{0,1,1} \\ m_{0,0,1} & m_{1,0,1} & m_{0,1,1} & m_{0,0,2} \end{pmatrix} \begin{pmatrix} \beta_0 \\ \beta_1 \\ \beta_2 \\ \beta_3 \end{pmatrix} = \begin{pmatrix} 1 \\ 0 \\ 0 \\ 0 \end{pmatrix} \tag{21}$$

where the moment $m_{a,b,c}$ is calculated by:

$$m_{a,b,c} = \sum_{j=1}^{N_j} (x_j - x_p)^a (y_j - y_p)^b (z_j - z_p)^c g_p(x_j - x_p) V_j \tag{22}$$

which designates the moment of order "$a$" about the x-axis, order "$b$" about the y-axis, and order "$c$" about the z-axis. Eq. (21) implies that the distributed force on fluid cells should not pose any torque about the particle center to be consistent with the original force, requiring all three torque components to be zero. In addition, the zeroth-order moment should equal unity due to the normalization or conservativity property.

In the 2nd-order RKPM (RKPM2), the corrector function is assumed as:

$$C_p(x) = \beta_0 + \beta_1 x + \beta_2 y + \beta_3 z + \beta_4 xy + \beta_5 yz + \beta_6 xz + \beta_7 x^2 + \beta_8 y^2 + \beta_9 z^2 \tag{23}$$

where the 10 coefficients $\beta_0$ to $\beta_9$ are chosen in such a way that the kernel satisfies 6 second-order consistency constraints in addition to the 4 first-order moments. The final system of equations for determining the coefficients for the kernel at each particle location is:

$$\begin{pmatrix} m_{0,0,0} & m_{1,0,0} & m_{0,1,0} & m_{0,0,1} & m_{1,1,0} & m_{0,1,1} & m_{1,0,1} & m_{2,0,0} & m_{0,2,0} & m_{0,0,2} \\ m_{1,0,0} & m_{2,0,0} & m_{1,1,0} & m_{1,0,1} & m_{2,1,0} & m_{1,1,1} & m_{2,0,1} & m_{3,0,0} & m_{1,2,0} & m_{1,0,2} \\ m_{0,1,0} & m_{1,1,0} & m_{0,2,0} & m_{0,1,1} & m_{1,2,0} & m_{0,2,1} & m_{1,1,1} & m_{2,1,1} & m_{0,3,0} & m_{0,1,2} \\ m_{0,0,1} & m_{1,0,1} & m_{0,1,1} & m_{0,0,2} & m_{1,1,1} & m_{0,1,2} & m_{1,0,2} & m_{2,0,1} & m_{0,2,1} & m_{0,0,3} \\ m_{1,1,0} & m_{2,1,0} & m_{1,2,0} & m_{1,1,1} & m_{2,2,0} & m_{1,2,1} & m_{2,1,1} & m_{3,1,0} & m_{1,3,0} & m_{1,1,2} \\ m_{0,1,1} & m_{1,1,1} & m_{0,2,1} & m_{0,1,2} & m_{1,2,1} & m_{0,2,2} & m_{1,1,2} & m_{2,1,1} & m_{0,3,1} & m_{0,1,3} \\ m_{1,0,1} & m_{2,0,1} & m_{1,1,1} & m_{1,0,2} & m_{2,1,1} & m_{1,1,2} & m_{2,0,2} & m_{3,0,1} & m_{1,2,1} & m_{1,0,3} \\ m_{2,0,0} & m_{3,0,0} & m_{2,1,0} & m_{2,0,1} & m_{3,1,0} & m_{2,1,1} & m_{3,0,1} & m_{4,0,0} & m_{2,2,0} & m_{2,0,2} \\ m_{0,2,0} & m_{1,2,0} & m_{0,3,0} & m_{0,2,1} & m_{1,3,0} & m_{0,3,1} & m_{1,2,1} & m_{2,2,0} & m_{0,4,0} & m_{0,2,2} \\ m_{0,0,2} & m_{1,0,2} & m_{0,1,2} & m_{0,0,3} & m_{1,1,2} & m_{0,1,3} & m_{1,0,3} & m_{2,0,2} & m_{0,2,2} & m_{0,0,4} \end{pmatrix} \begin{pmatrix} \beta_0 \\ \beta_1 \\ \beta_2 \\ \beta_3 \\ \beta_4 \\ \beta_5 \\ \beta_6 \\ \beta_7 \\ \beta_8 \\ \beta_9 \end{pmatrix} = \begin{pmatrix} 1 \\ 0 \\ 0 \\ 0 \\ 0 \\ 0 \\ 0 \\ 0 \\ 0 \\ 0 \end{pmatrix} \quad (24)$$

where the moment $m_{a,b,c}$ is calculated by Eq. (22).

Algorithm 4 of appendix A reports the steps of the 1$^{st}$-order RKPM. In the first step, $m_{a,b,c}$ is calculated by Eq. (22) as shown in lines 4 to 7. Then, the moment matrix in Eq. (21) is constructed in line 8. If the matrix is not well-conditioned ($RCond < 1e-3$), the PCM is substituted, as explained in lines 9 to 11. By a well-conditioned matrix, $\beta_i$ coefficients are obtained by the matrix inversion which is detailed in section 4. Finally, each cell's contribution $\phi_{pj}$ is computed using the modified kernel function as demonstrated in Eqs. (16)-(20) and lines 12 to 17. The algorithm of the 2$^{nd}$-order RKPM is similar, substituting Eq. (20) with (23) and (21) with (24). The determination of the cells within the ID of a particle is done with Algorithm 8.

*3.6. Reproducing Kernel Particle with Torque Method (RKPTM)*

In section 2.1, only the linear motion of particles was accounted for, and their rotational motion and its interaction with the continuous phase were neglected. For considering the rotational motion of particles, in the cases where the particle rotational velocity is significant or surface torques are dominant, the particle angular velocity, $\boldsymbol{\omega}_p$, should be tracked in addition to the particle velocity and location mentioned in section 2.1. Assuming dominant drag torque, the angular velocity equation can be written as:

$$I_p \frac{d\boldsymbol{\omega}_p}{dt} = \boldsymbol{T}_p; \quad \boldsymbol{T}_p = -\frac{C_R}{64}\rho_c d_p^5 |\boldsymbol{\omega}_r|\boldsymbol{\omega}_r \tag{25}$$

where $I_p = 0.1 m_p d_p^5$ is the particle moment of inertia, $\boldsymbol{\omega}_r = \boldsymbol{\omega}_p - 0.5(\boldsymbol{\nabla} \times \boldsymbol{u}_c)$, and the coefficient $C_R$ is given by the correlation derived from the direct numerical simulations by Dennis et al. [40] as:

$$C_R = \begin{cases} \dfrac{64\pi}{Re_R} & ; Re_R < 32 \\ \dfrac{12.9}{Re_R^{0.5}} + \dfrac{128.4}{Re_R} & ; 32 < Re_R < 1000 \end{cases} \tag{26}$$

and the rotational Reynolds number is defined by:

$$Re_R = \frac{d_p^2 |\boldsymbol{\omega}_r|}{v_c} \tag{27}$$

To account for the fluid-particle interaction torque in the two-way coupling source term of the continuous-phase momentum equation, the RKPM approach of section 3.5 can be resorted to with the difference that the first-order moments should equal the torque, $\boldsymbol{T}_p$, components rather than zero. In other words, the force $\boldsymbol{F}_p$ on the particle center should be spread over the Eulerian cells in such a manner that the resultant torque about the particle center equals $\boldsymbol{T}_p$. For this purpose, the force $\boldsymbol{F}_p$ is decomposed into two parts:

$$\boldsymbol{F}_p = \boldsymbol{F}_{p,\parallel} + \boldsymbol{F}_{p,\perp} \tag{28}$$

where ∥ and ⊥ indicate the direction parallel and perpendicular to $\boldsymbol{T}_p$, respectively.

$$\boldsymbol{F}_{p,\parallel} = (\boldsymbol{F}_p \cdot \boldsymbol{T}_p)\frac{\boldsymbol{T}_p}{|\boldsymbol{T}_p|^2}, \quad \boldsymbol{F}_{p,\perp} = \boldsymbol{F}_p - \boldsymbol{F}_{p,\parallel} \tag{29}$$

The force $F_{p,\|}$ is spread over the Eulerian grid by the standard RKPM, introduced in section 3.5, which exerts no torque about the particle center. The force $F_{p,\perp}$ and the torque $T_p$ about the particle center, $x_p$, is equivalent to the torque-free force $F_{p,\perp}$ exerted on the point $x_{p*}$ where $(x_{p*} - x_p) \times F_{p,\perp} = T_p$. It can be shown that:

$$x_{p*} = x_p + \frac{F_{p,\perp} \times T_p}{|F_{p,\perp}|^2} \tag{30}$$

Therefore, it is sufficient to spread $F_{p,\perp}$ in the region of influence of particle $p$ in such a way that the first-order moments about $x_{p*}$, rather than $x_p$, equal zero. For this purpose, the modified general moment is defined here by:

$$\widetilde{m_{a,b,c}} = \sum_{j=1}^{N_j} (x_j - x_{p*})^a (y_j - y_{p*})^b (z_j - z_{p*})^c \tilde{g}_p(x_j - x_p) V_j \tag{31}$$

For RKPM1, the consistency criteria are:

$$\begin{pmatrix} \widetilde{m_{0,0,0}} \\ \widetilde{m_{1,0,0}} \\ \widetilde{m_{0,1,0}} \\ \widetilde{m_{0,0,1}} \end{pmatrix} = \begin{pmatrix} 1 \\ 0 \\ 0 \\ 0 \end{pmatrix} \tag{32}$$

Substituting from Eqs. (19), (20), and (31) into (32) yields the following system of linear equations for the coefficients $\beta_0$ to $\beta_3$:

$$\begin{pmatrix} m_{0,0,0} & m_{0,0,0,1x} & m_{0,0,0,1y} & m_{0,0,0,1z} \\ m_{1,0,0} & m_{1,0,0,1x} & m_{1,0,0,1y} & m_{1,0,0,1z} \\ m_{0,1,0} & m_{0,1,0,1x} & m_{0,1,0,1y} & m_{0,1,0,1z} \\ m_{0,0,1} & m_{0,0,1,1x} & m_{0,0,1,1y} & m_{0,0,1,1z} \end{pmatrix} \begin{pmatrix} \beta_0 \\ \beta_1 \\ \beta_2 \\ \beta_3 \end{pmatrix} = \begin{pmatrix} 1 \\ 0 \\ 0 \\ 0 \end{pmatrix} \tag{33}$$

where $m_{a,b,c}$ is calculated by Eq. (22) and $m_{a,b,c,nx,my,qz}$ is computed by:

$$m_{a,b,c,nx,my,qz} = \sum_{j=1}^{N_j} (x_j - x_{p*})^a (y_j - y_{p*})^b (z_j - z_{p*})^c (x_j - x_p)^n (y_j - y_p)^m (z_j - z_p)^q g_p(x_j - x_p) V_j \tag{34}$$

We named this method as RKPTM1. The steps of this model are fully described in Algorithm 5. In this algorithm, first, the force $F_{p,\parallel}$ parallel to the applied torque is obtained and distributed using RKPM1 as described in lines 4 and 5. In order to distribute the perpendicular force $F_{p,\perp}$ and torque $T_p$ effect by an equivalent torque-free force, the new point of action, $x_{p*}$, is estimated in line 6. The components of the moment matrix in Eq. (33), $m_{a,b,c}$ and $m_{a,b,c,nx,my,qz}$, are calculated by Eqs. (22) and (34) in lines 8 to 11. Then, $\beta_i$ coefficients are computed by solving Eq. (33) which is noted in line 16. The rest of the algorithm for $\Phi_p = F_{p,\perp}$ is identical to RKPM1.

*3.7. Diffusion Two-Step Method (DTSM)*

Known as the diffusion-based model, Capecelatro and Desjardins [23] proposed the DTSM in 2013 which was later improved and made more efficient by Sun and Xiao [14] to include the PCM as the first step. The algorithm is applied in two steps: In the first step, a simple CG model, e.g., PCM, filters out data, and in the second step, the field obtained in the previous step is spread out by the solution of a diffusion equation using an adjustable diffusion coefficient. Consequently, two independent filtering steps are performed to attain the desired smoothness. The equation for the second stage is given by:

$$\frac{\partial \phi}{\partial \tau} = D_f \nabla^2 \phi, \quad 0 < \tau < 1 \tag{35}$$

where the diffusion coefficient, $D_f$, can be connected to the bandwidth, $b$, of a GKM CG by $D_f = b^2/4$ [14, 31]. $\phi$ is evolved in time by Eq. (35) until $\tau = 1$ starting from the initial condition obtained at the end of the first step and zero gradient or no-flux boundary conditions.

While DTSM is computationally much cheaper than GKM and RKPM methods for monodisperse flows; for polydisperse flows, the challenge is to determine the diffusion coefficient $D_f$ on the Eulerian grid. The most straightforward and accurate approach, based on the superposition principle, is to determine $D_f$ and solve Eq. (35) for each particle separately, which is called here the "reference" approach. However, it needs solving $N_p$ (the total number of particles in the domain) diffusion equations per Eulerian time step which is computationally inefficient and intractable. Capecelatro and Desjardins [23] suggested using the largest particle diameter in the domain, $d_{\max}$, to determine $D_f$. This is called the "constant" diffusivity approach, here. Nevertheless, for highly polydisperse flows with a wide range of particle diameters, a constant diffusion coefficient to spread the effect of all particles would be questionable. In this research, two additional approaches, called "group" and "local" models, for estimating $D_f$ in polydisperse flows are evaluated. In the "group" approach, the particle size range is divided into $N_G$ sub-ranges $(d_g, d_{g+1})$, called bins or groups. In the first step of DTSM. the contribution of each particle to $\phi_{pj}$ is collected in the group to which the particle belongs. Then, one diffusion equation is solved per particle group with the diffusion coefficient based on the group representative diameter, equal to the mid value of the bin range. Finally, the coarse-grained field equals the sum of fields of all groups. In the "local" approach, a single diffusion equation is solved using a field of $D_f$ which is varying on the Eulerian grid and locally determined based on the average particle diameter, $d_{\text{ave}}$, at each Eulerian cell and the global minimum particle diameter, $d_{\min}$. Therefore, it needs the computation of $d_{\text{ave}}$ field during the simulation. table 1 summarizes all approaches discussed here.

DTSM model implementation is described in Algorithm 6 of Appendix A. In lines 1 to 5, the field obtained in the first step of the CG algorithm using PCM is stored in $\phi_{j,g}$ which is assigned to each group $g$. The number of particle size groups, *particleGroupList*, is chosen in line 6 based on the $D_f$ calculation method ($N_p$ for "reference", $N_g$ for "group", and 1 otherwise). In lines 6 to 11, the second step is carried out.

**Table 1** The approaches for estimating diffusion coefficient of DTSM in a polydisperse flow ($D_{f,i} = b_i^2/4$, $b_i = \delta_i/6$, and $\delta_i = R_p d_{\text{ref},i}$).

| Approach | Number of equations solved (per Eulerian time step) | $D_{f,i}(d_{\text{ref},i})$ |
|---|---|---|
| Reference | $N_p$ | $d_{\text{ref},p} = d_p; p = 1, \dots, N_p$ |
| Group | $N_G$ | $d_{\text{ref},g} = \dfrac{d_g + d_{g+1}}{2}; g = 1, \dots, N_G$ |
| Constant | 1 | $d_{\text{ref},1} = d_{\max}$ |
| Local | 1 | $d_{\text{ref},1} = \begin{cases} d_{\min}; \text{ if cell is void} \\ d_{\text{ave}}; \text{ otherwise} \end{cases}$ |

### 3.8. Boundary treatment

A coarse-grained field must be modified to account for boundary effects when a particle approaches the domain boundaries. The PCM and GDM do not need any wall treatment since the unity of the sum of weight coefficients is ensured at each step.

For GKM, the original algorithm has a minor change near an inlet and outlet by setting the renormalization coefficient, $a_p$, to unity rather than using Eq. (14). For the wall or symmetry boundaries, two approaches are possible. In the first approach, no change to the original algorithm is made and the renormalization coefficient automatically takes care of keeping the conservativity of CG. The second approach, which is also used here, is called the "virtual particle" approach which results in a more physically-sound source distribution [14]; however, at a cost of extra computational overhead. The renormalization approach is more robust and straightforward, but it completely disregards the physics outside the boundaries [20].

The virtual particle model considers different particle-wall scenarios. No boundary treatment is applied when the particle ID and the wall boundaries do not intersect, see scenario "a" in figure 3. In scenario "b", a virtual particle is created by reflecting the main particle in the wall boundary in the cases when the particle ID overlaps a single wall. When the particle effect zone overlaps with two or more boundary walls, as indicated in scenario "c" in figure 3, the use of the virtual particle technique becomes more complicated [20]. This is accomplished by employing the recursive technique, which is continued until all newly generated virtual particles coincide with the predetermined particles. The successive virtual particles' effects are summed up for computing the cell weights. In scenario "c" depicted in figure 3, virtual particles 2 and 3 are created by reflecting the real particle in the left and top boundary walls. Finally, the reflections of virtual particle 4 coincide with the predetermined particles 2 and 3; thus, the recursive procedure is terminated. The total weight of each cell is evaluated by summing the weight of actual particles and the weight contributions from all virtual particles. Here, an efficient yet approximate "virtual particle" algorithm for GKM is presented in Algorithm 7 of Appendix A. In this approach, for each wall boundary and each grid cell, a cell-center-to-wall distance vector field is computed and stored at the beginning of the simulation (Sub-Algorithm 1). During the simulation, Sub-Algorithm 2 is responsible for finding virtual particles of any real particle. If a real particle ID and a wall overlap (by the approximate criterion in lines 2 and 6), the virtual particle location, $x_{vp}$, is computed and returned by Sub-Algorithm 2 to the main algorithm to account for the extra weights. The virtual particle is appended to *VPList* in line 7. The rest of the algorithm is repeated recursively until no new virtual particle is detected.

Although RKPMs are also based on the ID concept, no explicit boundary treatment is necessary for them. In DTSM, the boundary treatment is enforced easily with a zero-flux boundary condition for the diffusion equation in the second step.

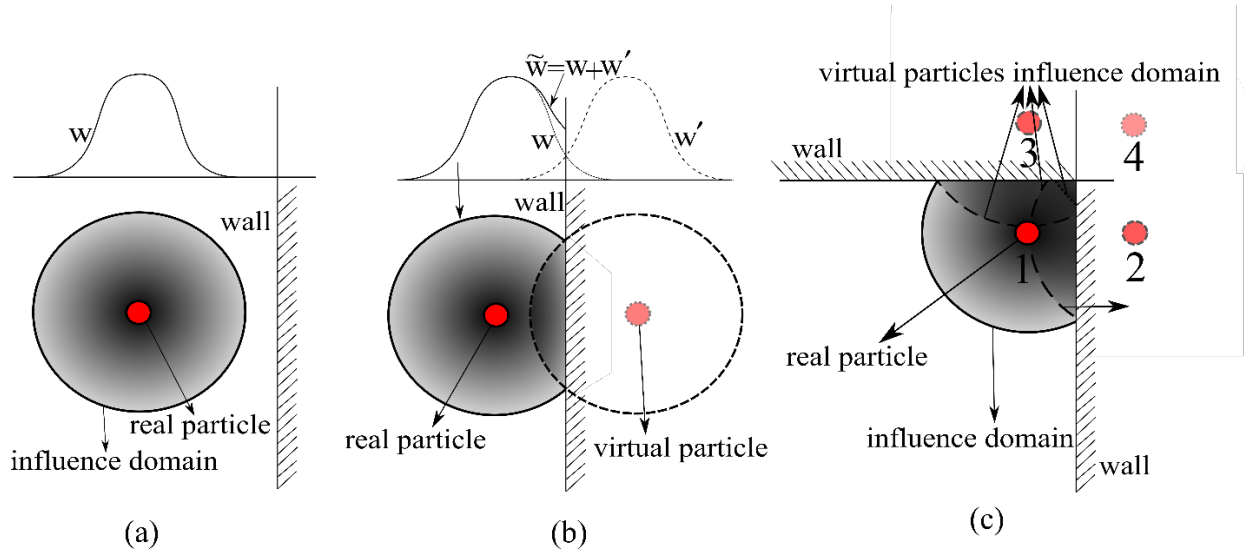

**Figure 3** The virtual particle boundary treatment: The particle domain of influence overlaps no boundary (a), a single boundary (b), and multiple boundaries (c).

## 4. Other numerical considerations

The open-source OpenFOAM CFD package (www.openfoam.org) version 2.3.0 was used for the solution of the governing equations. The Pressure-Implicit with Splitting of Operators (PISO) algorithm [41] is used for the pressure-velocity coupling in the Navier-Stokes equations. The standard second-order central differencing (Gauss linear) is adopted for the discretization of spatial derivatives [41]. The time derivatives are discretized by the implicit first-order (Euler) scheme [41]. The constant Eulerian time-step of $\Delta T_E = 4 \times 10^{-4}\ s$ is used in all test cases. For all variables, a normalized residual tolerance of $10^{-6}$ is set as the convergence criterion at each time step.

The face-to-face procedure [42] is used in OpenFOAM for particle tracking. In the Lagrangian particle tracking equations, the continuous-phase properties are interpolated to the particle locations. Here, the fluid velocity is interpolated using the "cellPoint" algorithm. In this technique, first, particle properties are computed on the Eulerian grid vertices using the linear interpolation scheme. Afterward, the properties at a location within the cell are estimated using the methodology introduced by Smith and Ebert [4] using the Barycentric coordinates. The analytical method [43] is chosen for the integration of the particle Lagrangian equations. The integration time step for each particle is determined independently by:

$$\Delta t_p = \frac{Co_L \sqrt[3]{V_c}}{|\boldsymbol{u}_p|} \tag{36}$$

where $Co_L$ is the Lagrangian Courant number which is chosen to be 0.3 in this study. In addition, as the volume fraction value approaches unity, the solution becomes unstable [31]. In this research to prevent numerical instabilities, the solid volume fraction is restricted to the upper bound of 0.9.

The OpenFOAM "Lagrangian" library, which deals with EL simulation, was extended to apply all CG models and related algorithms introduced in section 3. The OpenFOAM Lagrangian step is described in Algorithm 9 of Appendix A. According to this algorithm, particles are injected using OpenFOAM built-in injection models in line 2. In line 10, the local Lagrangian time step for particle $p$, $\Delta t_p$, is restricted by three criteria, including the Lagrangian Courant number, Eq. (36). Over $\Delta t_p$, the particle position and velocity are integrated through lines 11 to 15. Note that, in line 13, for a stronger coupling between the continuous and dispersed phases, the continuous-phase velocity is modified at each local Lagrangian time step ($\Delta t_p$) by the interphase force obtained in the previous local Lagrangian step as:

$$\boldsymbol{u}_c^{mod} = \boldsymbol{u}_c + \frac{\boldsymbol{F}_p}{\rho_c V_c} \tag{37}$$

The CG operation can be directly included after the overall Lagrangian step, by a call at line 21. Alternatively, the CG loop over all particles (line 1 of all CG algorithms) can be combined with the main loop in the "Lagrangian" step (line 3 of Algorithm 9), i.e., removing the loop over particles (line 1) in CG algorithms and place the call to CG algorithm in line 17, as used in this work. With this latter approach, the CG algorithm is performed for each particle once or more, depending on the number of local Lagrangian time steps in each Eulerian step. In this case, to keep the CG algorithm correct, it is sufficient to multiply the original $\Phi_p$ by the factor $\Delta t_p/\Delta T_E$ (line 17). This is equivalent to the run-time averaging used in references [2, 44]. Although this approach poses a computational overhead compared to the first one, the decreased statistical error of the two-way coupling source terms enhances the accuracy and robustness of the solver and makes it possible to use larger Eulerian time steps for the simulations. If the DTSM algorithm is invoked, its second step is performed after the main Lagrangian loop in line 21. The time step for the solution of each diffusion equation in this step, Eq. (35), is chosen as $\tau/30$ along with the normalized residual tolerance of $10^{-6}$ at each time step.

Here, for RKPM models, $\beta_i$ coefficients are calculated by solving the linear systems of equations using matrix inversion. For this purpose, Armadillo algebraic computing software library [45] was integrated into OpenFOAM. When the ratio of the particle diameter to cell length scale, $d_p/\Delta$, is small and the particle ID comprises few cells, i.e., a small ID radius, some coefficients in the moment coefficient matrix approach zero, which may cause singularities and solution divergence. This is more probable for the $2^{nd}$-order RKPM. To address this issue, the coefficient matrix reciprocal condition number [46] or "RCond", which is between 0 and 1, is calculated. The value of RCond is near one for a well-conditioned matrix and close to zero for an ill-conditioned situation. In other words, the lower the RCond, the closer the matrix to the singularity. To prevent the solution divergence in extreme conditions, if RCond is less than $10^{-3}$, the algorithm switches from RKPM to PCM. This does not pose a serious limitation on the present RKPM models since when $d_p/\Delta \ll 1$, PCM performs satisfactorily with a low computational cost. Therefore, the present RKPM models are, in fact, hybrid models which take advantage of pure PCM and RKPM CG algorithms at $d_p/\Delta \ll 1$ and $d_p/\Delta \geq 1$, respectively.

## 5. Results and discussion

The performance of CG algorithms is evaluated using four benchmarks. Here, 3D test cases with spherical particles are designed while particles are distributed over a plane to pose more challenge CG algorithms by more inhomogeneous coarse-grained fields. For quantitative assessment of the grid independency of CG algorithms, a measure is defined and denoted by $C_{m \to n}^{f}$ which shows the variation of a field $f$ due to the change of computational grid from a resolution $m$ to $n$ and is defined by:

$$C_{m \to n}^{f} = \frac{\sum_{i=1}^{N}|f_i^n - f_i^m|}{\sum_{i=1}^{N}|f_i^m|} \tag{38}$$

where $f_i^m$ and $f_i^n$ are the field values on grid $m$ (base grid) and $n$, respectively, at sampling location $i$, computed using the linear interpolation, and $N$ is the number of chosen sample points. The algorithms with lower $C_{m \to n}^{f}$ values are deemed of higher degrees of the grid independency. $N$ or the number of sampling points is 121 (corresponding to the finest grid in table 3) for all benchmarks in this study. Note also that, $C_{m \to n}^{f}$ is computed using the direct outputs of CG algorithms before imposing the *ad hoc* limiter to bound the maximum $\theta_s$ to 0.9.

### 5.1. The "falling particles" benchmark

In this 3D benchmark, four particles with the density of 1000 $kg/m^3$ are released from rest on the plane $x = -1.2\ m$ at $t = 0\ s$ in a cubic domain with the edge length of 3 $m$ centered at the origin of the coordinates system and filled with initially stagnant air (the continuous phase). The gravity is in the negative z-direction. The symmetry boundary condition is applied at the top and bottom boundaries, and the no-slip condition is assigned to the side walls. The initial particles' positions and diameters are reported in table 2 and figure 4. The fluid and particles interact through the drag force and solid volume fraction. The results are analyzed on the plane $x = -1.2\ m$. For computing the grid independence measure by Eq. (38), 121 samples, uniformly distributed on the sampling line (shown in figure 4), are taken at $t = 0.0012\ s$.

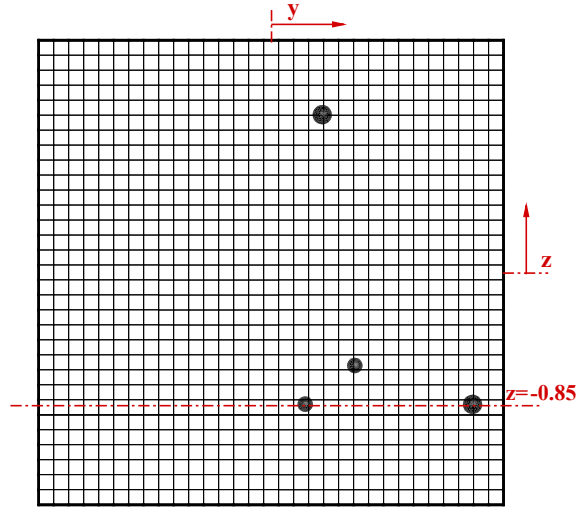

**Figure 4** Benchmark 1 (the "falling particles"): Particles are shown at their initial positions, and the sampling line ($x = -1.2\,, z = -0.85$) with the red dash-dotted line.

**Table 2** Benchmark 1: The particles' initial position and diameter.

| Particle ID | Particle position ($m$) | Particle diameter ($m$) |
|---|---|---|
| 1 | (-1.2 0.22 -0.85) | 0.1 |
| 2 | (-1.2 0.54 -0.6) | 0.1 |
| 3 | (-1.2 1.3 -0.85) | 0.125 |
| 4 | (-1.2 0.33 1.02) | 0.125 |

*5.1.1.Grid independency: The grid resolution*

In the first grid-independence test, the effects of grid resolution variation on the coarse-grained fields are examined. Uniform structured grids are employed for this test, and only the number of cells in each direction ($N_{x,y,z}$) is changed. table 3 summarizes the grid sizes and the ratio of the particle diameter to cell length scale, $d_p/\Delta$, for each grid. $N_{cell}$ is the total number of cells of each grid. The simulations are conducted using different CG algorithms. In figure 5, the contours of the solid volume fraction, $\theta_s$, are plotted over the plane $x = -1.2\ m$ at $t = 0.0012\ s$. For a better demonstration, the animated contours of the solid volume fraction are provided for all CG models in "video1" in the supplementary materials. In figure 5, the extent and trend of the particle presence effect on the solid volume fraction distribution over the grid cells can be compared for different CG algorithms. With the PCM, particle spreading is restricted to the host cell. This results in a large peak of $\theta_s$ at cells where a particle center resides. In fact, the predicted $\theta_s$ by PCM was larger than 1 and was limited to the maximum allowable value of 0.9 by the *ad hoc* boundness condition to prevent instability issues. The GDM algorithm mitigates this issue to some extent by distributing the particle presence effect, or volume, to the host neighbor cells. With other CGs, the particle effect is widely distributed in the domain of influence. Note that, compared to 2D examples considered in other studies [14, 47], in 3D cases, the particle variables spread across many more cells, and thus, the maximum solid volume fraction is significantly smaller than the one in 2D cases with the same ID diameter.

**Table 3** The grid resolution test: The specification of different uniform grids.

| Grid | $N_{x,y,z}$ | $N_{cell}$ | $N/N_F$ | $d_p/\Delta$ (benchmark 1) | (benchmark 2) | (benchmark 3) | (benchmark 4) |
|---|---|---|---|---|---|---|---|
| A | 21 | 9261 | 0.0005 | 0.7-0.875 | 0.42 | - | - |
| B | 41 | 68 921 | 0.039 | 1.37-1.71 | 0.82 | - | - |
| C | 61 | 226 981 | 0.128 | 2.03-2.54 | 1.22 | - | 2.54 |
| D | 81 | 531 441 | 0.299 | 2.70-3.37 | 1.62 | - | - |
| E | 101 | 1 030 301 | 0.581 | 3.37-4.20 | 2.02 | - | - |
| F | 121 | 1 771 561 | 1 | 4.03-5.04 | 2.42 | 1.21-12.1 | - |

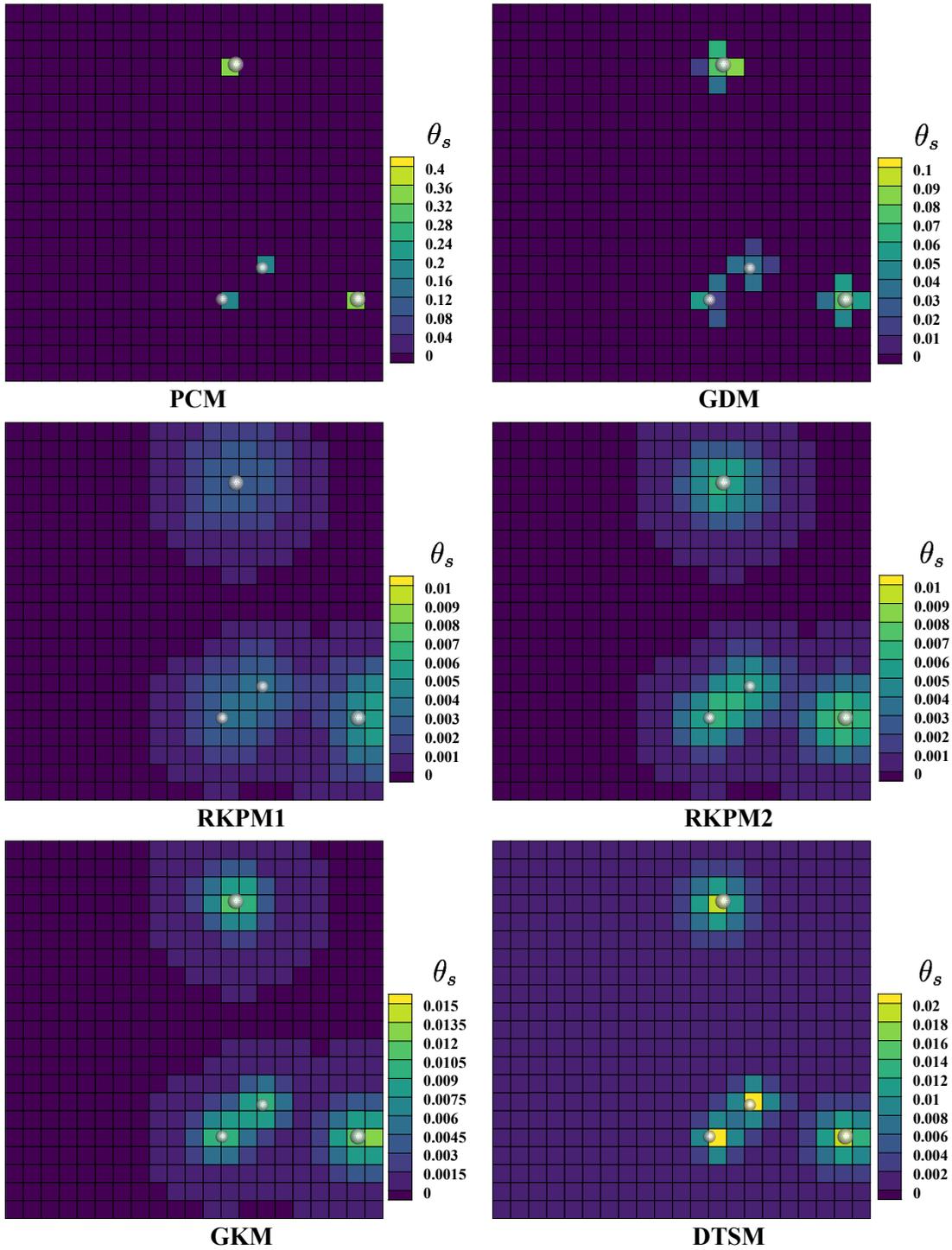

**Figure 5** Benchmark 1: The contours of solid volume fraction on the plane $x = -1.2\ m$ at $t = 0.0012\ s$ for different CG algorithms on grid A (table 3).

In the grid resolution test (table 3), the particle diameter to cell length scale ratio ($d_p/\Delta$) varies from 0.7 for the coarsest grid (grid A) to about 5.0 for the finest grid (grid F). The $\theta_s$ profiles along the sampling line for different CG algorithms and four different grids are illustrated in figure 6. As can be observed, the PCM and GDM are far from being grid independent. In addition, there exists high volume fraction gradients and local peaks at the particle positions in their results. According to figure 6a and b, for all grids but the coarsest one, $\theta_s$ predictions by PCM and GDM exceed the limit of 0.9 and are cut by the maximum volume fraction condition to maintain stability. The actual predicted values of $\theta_s$ by PCM at the particle 3 (table 2) location, without the stability restriction, were 0.35 and 67 on grid A ($N_{x,y,z} = 21$) and F ($N_{x,y,z} = 121$), respectively. Nevertheless, this kind of *ad hoc* criteria cannot necessarily provide the solution stability, since only $\theta_s$ can be limited by a predefined threshold while the interphase coupling force ($\boldsymbol{f}_{FP}$ in Eq. (7)) grows unboundedly by increasing $d_p/\Delta$, which destabilizes the solution or decreases the accuracy in the best scenario. Moreover, this simple stability criterion loses the conservativity property of the CG.

The rest of the models show reasonable levels of grid independence as the grid size reduces. figure 6c illustrates the results of the RKPM1 algorithm. As can be observed, this model yields smooth distributions of particle volume fraction with a high level of grid independence. The red dashed lines in figure 6 for RKPM and GKM indicate the particle ID borders. Based on these borders, the grid cells with a center of $y > 0.82$ are only affected by particle 3 (table 2). The cells within $0.55 < y < 0.82$ receive contributions from both particles 1 and 3, and cells with the center within $-0.38 < y < 0.55$ are influenced only by particle 1 while other cells ($y < -0.38$) along the sampling line do not receive contributions from any particle.

The RKPM2 model in figure 6d results in a relatively more compact solid volume fraction distributions and larger peaks compared to RKPM1. More importantly, there is an oscillation and a local maximum in the distribution near the border of the ID which is not present in the RKPM1 profile. The form of the kernel function is the major source of this difference between RKPM1 and 2. The computed $\beta_i$ coefficients of RKPM1, i.e., Eq. (21), resulted in a broader profile with a lower peak at the particle center in comparison to RKPM2. In addition, the non-linear nature of RKPM2 kernel modifier function, i.e., Eq. (23), compared to the linear modifier in RKPM1, i.e., Eq. (21), brings about the aforementioned oscillating profiles of RKPM2 near the ID boundaries.

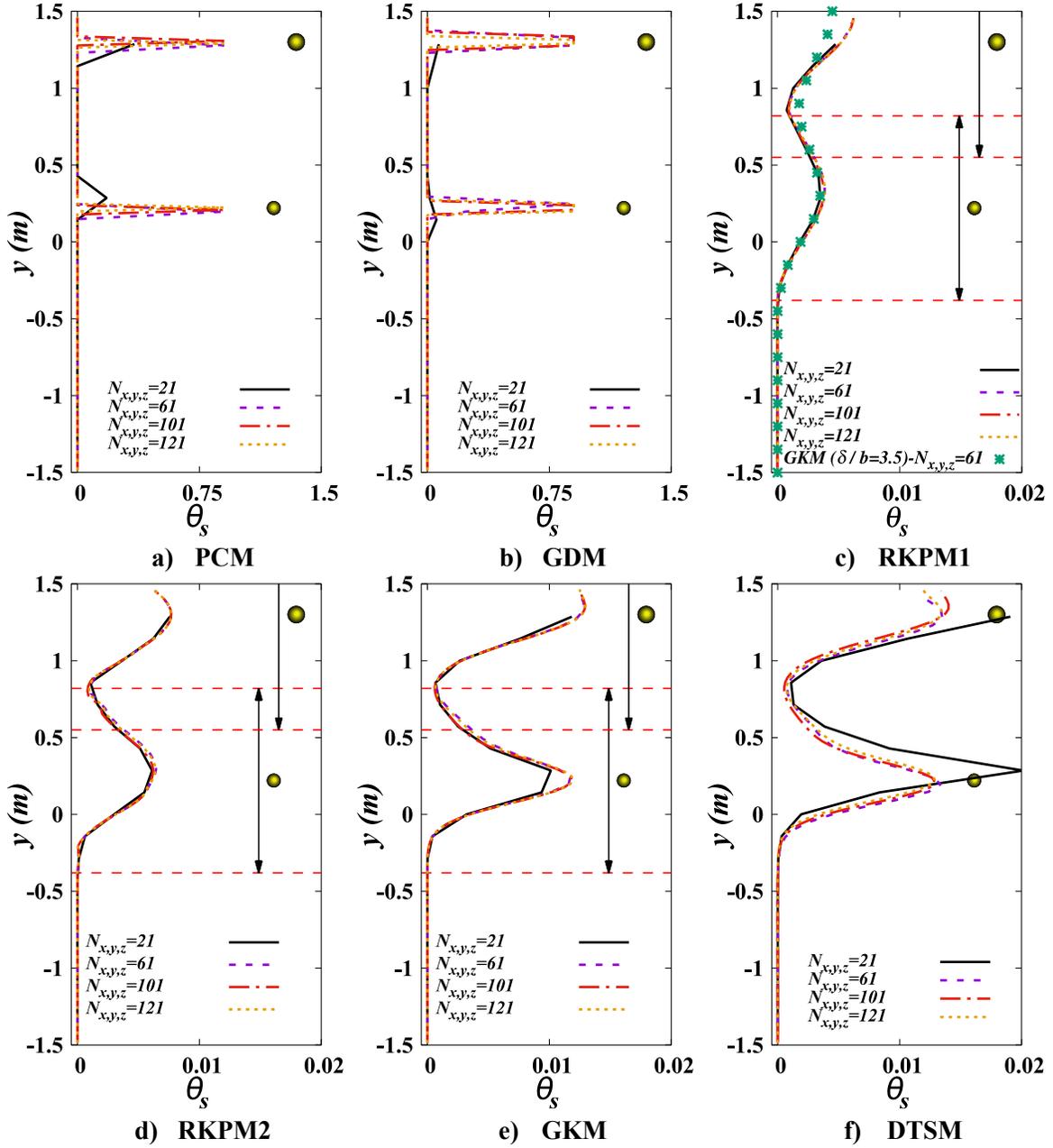

**Figure 6** Benchmark 1: The solid volume fraction along the sampling line (see figure 4) at $t = 0.0012\ s$ for different grid resolutions and CG algorithms. Particle locations and domain of influence limits (red dashed lines) are also illustrated.

GKM results in figure 6e show the fine grid independence property of GKM. The asymmetry of the GKM profiles for the particle falling near the wall is evident in this figure. This is due to the effect of the "virtual particle" approach for boundary treatment in GKM. The larger peak volume fraction predicted by the GKM algorithm compared to the ones of RKPM1 stems from the parameters used to define the template functions of these models. To clarify this point, the Gaussian kernel function, Eq. (15), in GKM and Peskin's cosine kernel, Eq. (18), in RKPM are plotted against each other in figure 7. As can be seen, the Gaussian kernel with the default value of $\delta_p/b = 6$ [14] results in a much more compact support and larger peak compared to the cosine template function. Note that to reduce the difference between GKM and RKPM, a different value $\delta_p/b$ can be chosen. It should be noted that to preserve the conservativity property of the algorithm, the Gaussian kernels shown in figure 7, are in fact the renormalized Gaussian kernels which differ by a factor of $a_p$ (Eq. (14)) from Eq. (15). We found that the best approximate to the cosine form with the Gaussian from is obtained choosing $\delta_p/b = 3.5$. As it is observed in figure 7, this kernel has the smallest overall deviation from the cosine kernel with a small jump at the ID borders in contrast to the cosine form which continuously approaches zero at the ID borders.

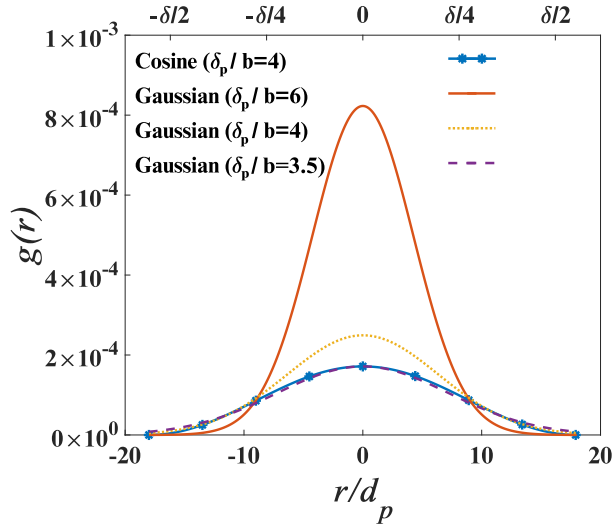

**Figure 7** The comparison of the Peskin's cosine template function and the Gaussian template function with different $\delta/b$ values. The diagram shows the 2D cut of the 3D isotropic kernels versus the (normalized) distance from the kernel center.

To compare the results of GKM using the modified Gaussian kernel ($\delta_p/b = 3.5$) with the ones of RKPM1, the results of the GKM with $\delta_p/b = 3.5$ is also included in figure 6d. As can be seen, the solid volume fraction distributions with both kernel types are close to each other near particle 1 in the middle of the domain, still, there is a deviation in the solid fraction distributions near the walls which is attributed to the distinct methods of boundary treatment used in these two CG algorithms.

The DTSM results in figure 6f show a reasonable grid-independence property, however, at higher grid resolutions compared to RKPM and GKM algorithms. This is due to the fact that the first step of the DTSM algorithm is PCM which is grid-dependent itself. Based on the proof given by Sun and Xiao [48], the result of the second step, i.e., the solution of the diffusion equation, is equivalent to the one of the grid-independent GKM when the initial condition, which is provided in the first step, is the superposition of multiple Dirac delta functions. This is the case for PCM when the grid size approaches small values. As a result, DTSM can provide grid-independent solutions at $d_p/\Delta \gg 1$ or fine grid resolutions. In addition, the nearly identical results of GKM and DTSM (at large $d_p/\Delta$) can be seen in figure 6 which is compatible with the theoretical proof given by Sun and Xiao [48]. A more quantitative discussion on the performance of different models will be made in section 5.4.

*5.1.2. Grid independency: The grid non-uniformity and stretching*

In practical CFD problems, the use of non-uniform grids is unavoidable. For wall-bounded flows, the grid spacing is usually changed in the wall-normal direction which results in stretched grids with high aspect ratios, especially near the walls. To check the performance of different algorithms on stretched grids, two scenarios are considered here. In the first one, the grid resolution is varied only in the y-direction (1D-stretched grid in figure 8a). The maximum cell-size ratio is 5, i.e., the grid expansion factor of 1.13, which results in the maximum aspect ratio of 2.55 and the maximum cell-volume ratio of 5 for this grid. In the second scenario, the stretching is applied in all 3 directions (3D-stretched grid in figure 8b). For this case, the maximum aspect ratio is 5.21 and the maximum cell-volume ratio is 125. In order to assess the model sensitivity to mesh stretching, the uniform grid with the same number of cells, i.e., grid "C" of table 4, is chosen as the reference grid.

figure 9 shows $\theta_s$ predictions by RKPM1, RKPM2, GKM, and DTSM on different grids. Similar to the previous test, PCM and GDM show high degrees of grid sensitivity and their results are not shown for the present test for the sake of brevity. According to figure 9, RKPM variants and GKM show high levels of insensitivity to the grid stretching. The grid-independence property of DTSM is also acceptable; however, some grid dependency is observed for DTSM near the right wall boundary. This may be due to the discretization error of the solution of the diffusion equation on the highly stretched grid with a large aspect ratio near the walls. The great performance of RKPM models under the grid stretching test reflects the success of the $1^{st}$ and $2^{nd}$-order corrector functions in RKPMs under this kind of grid non-uniformity. These models show excellent grid independency as the ratio of $\delta_p/\Delta$ increases. This is because the assumption of the constant kernel function value over each grid cell used in these models becomes more accurate by increasing $\delta_p/\Delta$.

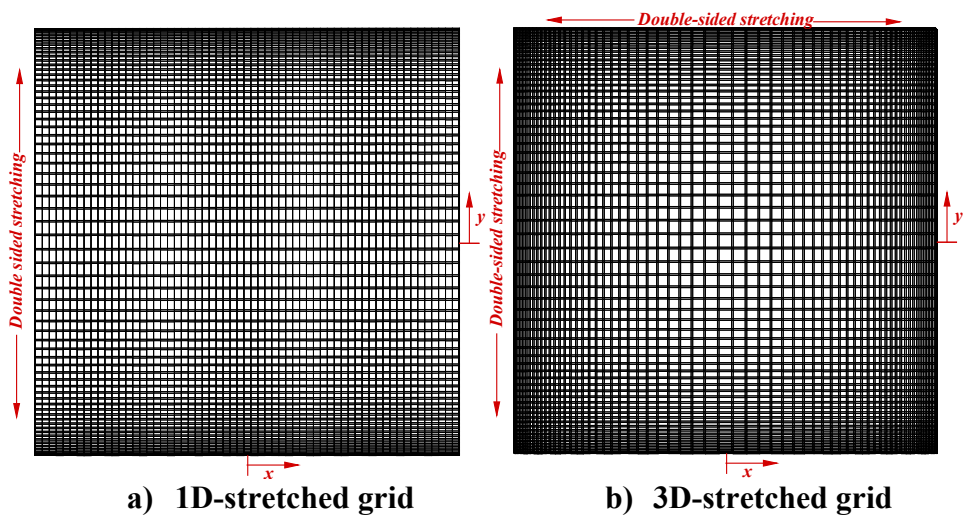

a) 1D-stretched grid    b) 3D-stretched grid

**Figure 8** The mesh cross section view for $N_{x,y,z} = 61$ for two different types of grid streching.

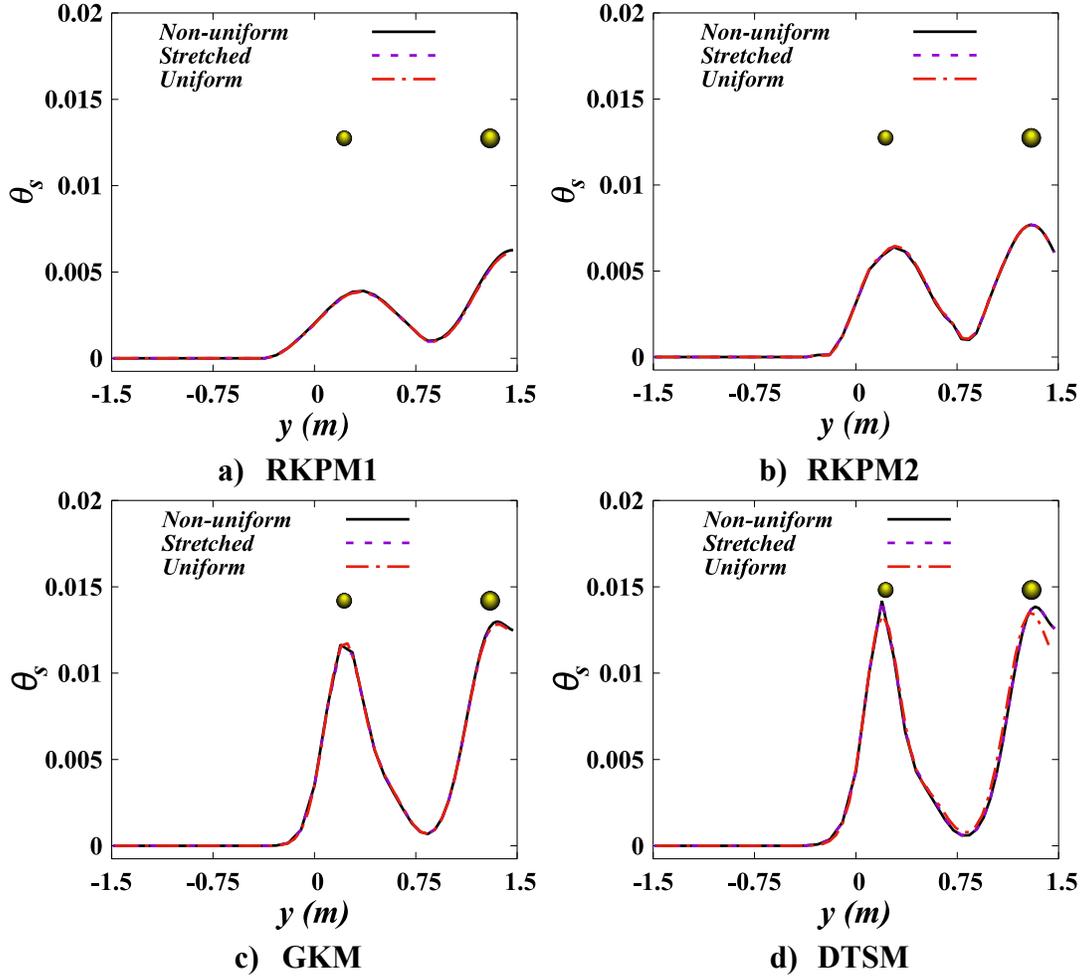

**Figure 9** Benchmark 1: The solid volume fraction along the sampling line (see figure 4) at $t = 0.0012\ s$ for different types of grid stretching and CG algorithms.

*5.1.3. Grid independency: The grid type and non-orthogonality*

The CFD of problems with complex geometries often necessitates the use of unstructured grids. Here, the performance of the CG algorithms on different unstructured grids with different degrees of distortion and non-orthogonality is evaluated. For this purpose, the solution on grid "C" of table 4 is considered the reference solution. The properties of this grid are given in the first row of table 4. In this table, the specification of 3 other unstructured grids with tetrahedral cells is also reported. The unstructured grids, shown in figure 10, are devised in such a manner that their total number of cells, $N_{cell}$, are close to the one of the structured grid (within 2% deviation). $N_{LF}$ and $N_{RF}$ are the number of uniform divisions in each direction on the left and right boundary of the domain, respectively. The ratio of $N_{RF}/N_{LF} = n$ is denoted by $r:n$ for each unstructured grid. The four boundary edges in the y-direction, connecting the left and right faces of the cubic domain, is divided with a constant cell expansion factor, given the cell sizes on the right and left boundaries. Knowing the aforementioned divisions of the 12 edges of the cubic domain, an unstructured grid with tetrahedral cells is generated for each case. According to table 4, the maximum skewness or non-orthogonality increases for these grids as $N_{RF}/N_{LF}$ grows.

Table 4 The grid specification for grid type and non-orthogonality test.

| $N_{LF}$ | $N_{RF}$ | $N_{cell}$ | Maximum skewness | Maximum aspect ratio | Grid ID |
|---|---|---|---|---|---|
| 61 | 61 | 226981 | 0 | 1.0 | Structured (grid C) |
| 31 | 31 | 224834 | 0.61 | 5.72 | Unstructured (r:1) |
| 20 | 40 | 228959 | 0.63 | 7.16 | Unstructured (r:2) |
| 11 | 44 | 225850 | 0.81 | 6.83 | Unstructured (r:4) |

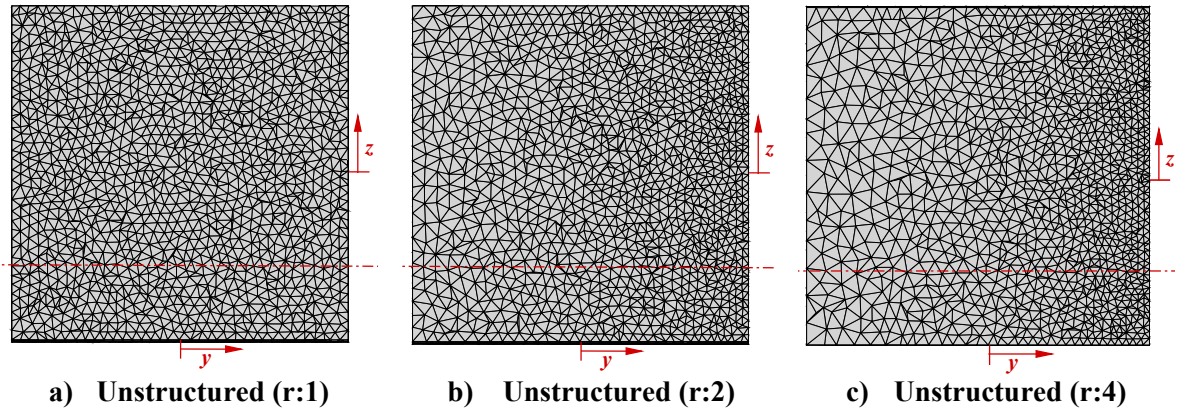

a) Unstructured (r:1)  b) Unstructured (r:2)  c) Unstructured (r:4)

**Figure 10** The y-z cross-section of the unstructured grids introduced in table 4.

The predictions of different models under the grid type and non-orthogonality test are demonstrated in figure 11. GKM and RKPM variants show high levels of grid-independence property in this test too. Nevertheless, there is some level of high-frequency oscillations in the predictions of these three CG algorithms. This manifests that the grid non-orthogonality test poses more challenges to CG algorithms. The level of fluctuations in RKPM1 is lower than RKPM2 and GKM which suggests that the 1$^{st}$-order corrector function of RKPM1 performs better for the skewed grids. Similar to the grid stretching test, DTSM shows a higher level of grid sensitivity compared to GKM and RKPMs. This is primarily attributed to the sensitivity of the algorithm to the initial condition provided through the first PCM step and, then, to the discretization error of the diffusion equation. The grid-sensitivity of DTSM can also be observed in the results reported in other studies; see e.g., reference [49].

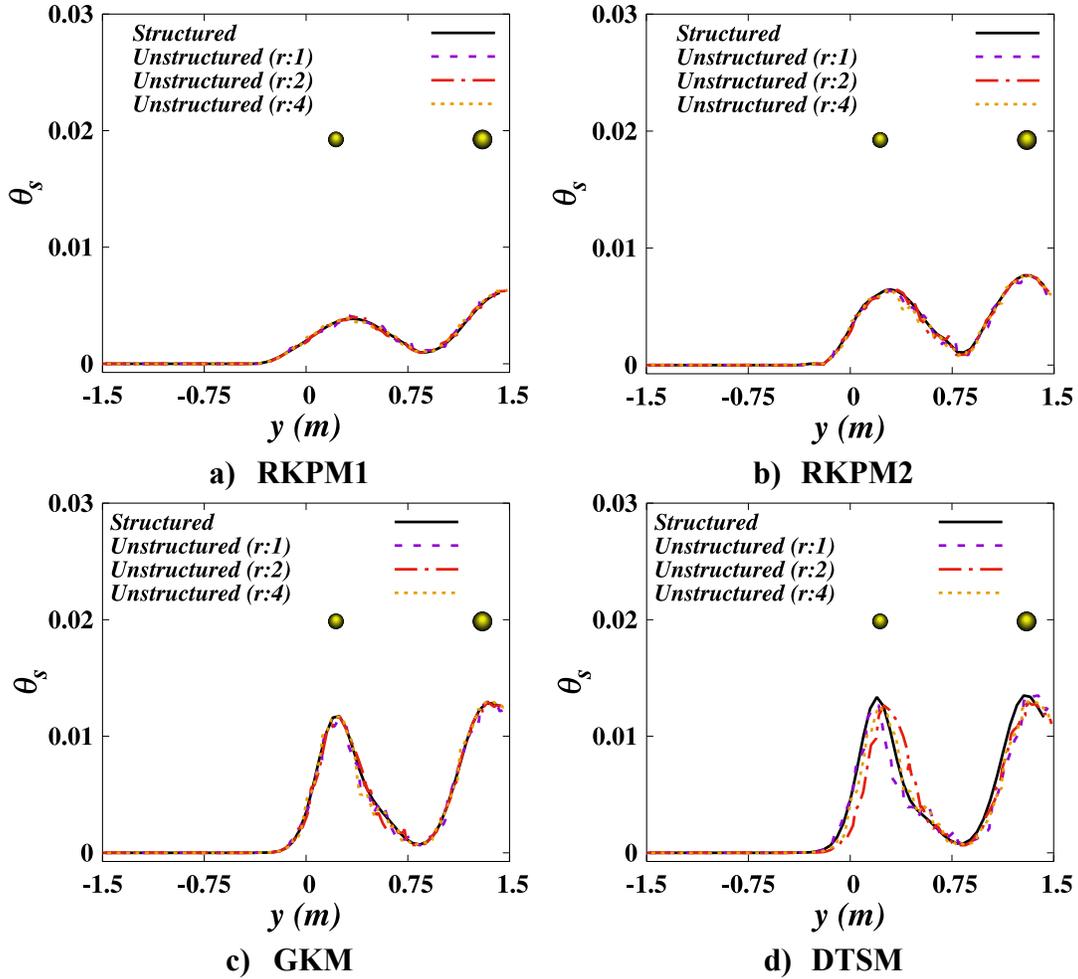

**Figure 11** Benchmark 1: The solid volume fraction along the sampling line (see figure 4) at $t =$ 0.0012 $s$ for different grid non-orthogonality levels (see table 4) and CG algorithms.

*5.2. The "random particle distribution" benchmark*

The 3D "random particle distribution" benchmark is devised to assess the performance of CG algorithms under high solid volume fraction values (dense regimes) with a $d_p/\Delta$ ratio from 0.42 to 2.42. In this test case, 5000 particles with a diameter of 4 cm and density of 1000 $kg/m^3$, which are uniformly distributed on the plane $z = 0$ at $t = 0$ in a cubic domain with an edge length of 2 m, are released from rest. The maximum overlaps of $0.4d_p$ and $0.2d_p$ are allowed for the particle-particle and particle-wall initial arrangement, respectively. The domain is centered at the origin of the coordinates system and filled with initially stagnant air. Gravity acts in the negative z-direction. The symmetry boundary condition is applied at the top and bottom boundaries while the no-slip and the simple reflection are adopted for the continuous and dispersed phase, respectively, at the four side boundaries. figure 12 demonstrates the cross-section of the computational domain and initial arrangement of particles. The results are analyzed over the plane $z = 0$. For computing the grid independence measure by Eq. (38), 121 samples, uniformly distributed on the sampling line (shown in figure 12), are taken at $t = 0.0012\ s$.

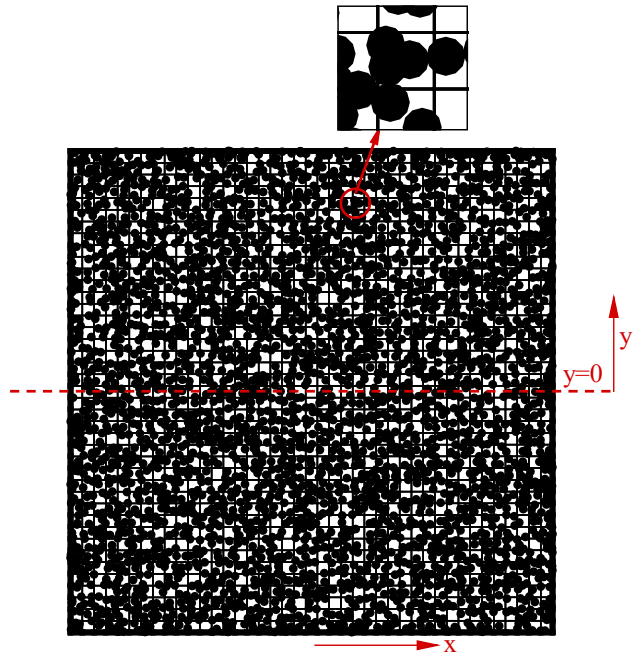

**Figure 12** Benchmark 2 (the "random particle distribution" benchmark): The $z = 0$ cross-section of the computational domain, particle initial distribution, and grid (grid B of table 3). Particles are depicted with the size scaling of 1:1.25 for a better demonstration. The sampling line ($y = 0, z = 0$) is plotted with the red dashed line.

*5.2.1. Grid independency: The grid resolution test*

For the grid resolution test in this section, the grids introduced in table 3 are adopted. According to this table, in benchmark 2, $d_p/\Delta$ varies between 0.42 to 1.62 for grids A to F. The predicted $\theta_s$ profiles by different CG models at the sampling line are compared in figure 13. The predictions by PCM are noisy profiles with large fluctuations truncated to the specified threshold of $\theta_s = 0.9$ except for the coarsest grid where $d_p/\Delta < 0.5$. In case, the threshold $\theta_s = 0.9$ is not enforced on the results, the PCM profile peaks at $\theta_s = 4.71$ for this benchmark. To see the deviation of PCM predictions from the physical maximum possible peak value, the local maximum volume fraction at the packing limit can be considered. For 2D problems (with 2D cylindrical particles) and non-overlapping particles, this value can be easily computed as 0.938 for $d_p/\Delta = 1$. For 3D problems and spherical particles when particles can overlap within a specified range, i.e., $0.4d_p$ in the present benchmark, the calculation of the value of the maximum possible local solid volume fraction is rather complicated. For this purpose, we used the python package developed by Strobl et al. [50]. Based on this calculation, the maximum possible value is $\theta_s = 2.078$ which is less than half the PCM peak value (4.71). Therefore, PCM does not show grid-independence characteristics with a peak value of $\theta_s$ which is about 130% larger than the maximum possible solid fraction. GDM predictions in figure 13b improve on the ones by PCM; The oscillations are smaller and the maximum possible value of 2.078 is not violated in the unconfined results. However, there is still a considerable level of grid dependency in the results and high solid volume fraction gradients in the oscillating profiles which can hinder the stability and convergence.

The $\theta_s$ predicted by RKPM variants and GKM in figure 13c-e show high levels of grid independence, the results on all grids except the coarsest one ($N_{x,y,z} = 21$) coincide with each other along the sampling line. RKPM results in the increase of the predicted volume fraction by approaching the walls like the effect of wall treatment in GKM described in section 3.8, though the wall boundary treatment is different for RKPMs. In contrast to GKM, RKPMs do not use an *ad hoc* wall treatment. The mechanism of wall treatment in RKMPs can be described as follows. Assuming particle "p" resides near a wall with wall-normal direction in the positive x-direction (x+ wall), the Eulerian grid cells in the particle ID are mainly located on one side of the particle in such a manner that the term $(x_j - x_p)$ in Eq. (22) is predominantly positive in the summation. As a result, the $m_{1,0,0}$ moment becomes a large positive number compared to the other first-order moments, $m_{0,1,0}$ and $m_{0,0,1}$, which are small due to the nearly symmetric distribution of cells in the y and z directions around the particle. Based on an order of magnitude analysis, it can be shown that:

$$m_{0,0,0} \gg m_{1,0,0} \gg m_{0,2,0}, m_{0,0,2}, m_{2,0,0} \gg m_{0,1,0}, m_{0,0,1} \gg \text{other moments} \tag{39}$$

Therefore, the solution of Eq. (21) satisfies:

$$\beta_0 > 0, \beta_1 < 0, \beta_3 \sim \beta_4 \sim 0 \tag{40}$$

and the corrector function, Eq. (20), is simplified to $C_p(x) = \beta_0 + \beta_1 x$. With $\beta_1 < 0$, the negative slope of $\theta_s$ in the x-direction near the x+ wall is justified.

The level of $\theta_s$ at the sampling line predicted by RKPM2 is higher than that of RKPM1 (see figure 13c and d). In addition, the oscillations in the RKPM2 profiles are more pronounced. These observations were also made for the first benchmark and are attributed to the non-linear corrector function used in the RKPM2 kernel function. It should be noted that as the grid coarsens and $d_p/\Delta$ reduces, fewer cells are embedded in each particle ID, particularly near the wall regions. This can bring about a nearly ($RCond \approx 1e - 3$) or entirely ($RCond < 1e - 3$) ill-conditioned coefficient matrix in RKPMs. For the present test case, this occurred solely for RKPM2 on the coarsest grid ($N_{x,y,z} = 21$ and $d_p/\Delta = 0.42$) and only for 11 particles among the total 5000 particles. For these particles, the hybrid RKPM2 switched to PCM as planned. While for RKPM1, the coefficient matrix was well-conditioned and no need for switching to PCM was reported in all benchmarks of the present study. The value of $\theta_s$ predicted by GKM in figure 13e is higher than the RKPM predictions. That is due to the more compact kernel shape in the standard GKM discussed in section 5.1.1. The deviation of GKM and RKPM1 can be minimized away from the walls by using the modified Gaussian kernel bandwidth described in section 5.1.1.

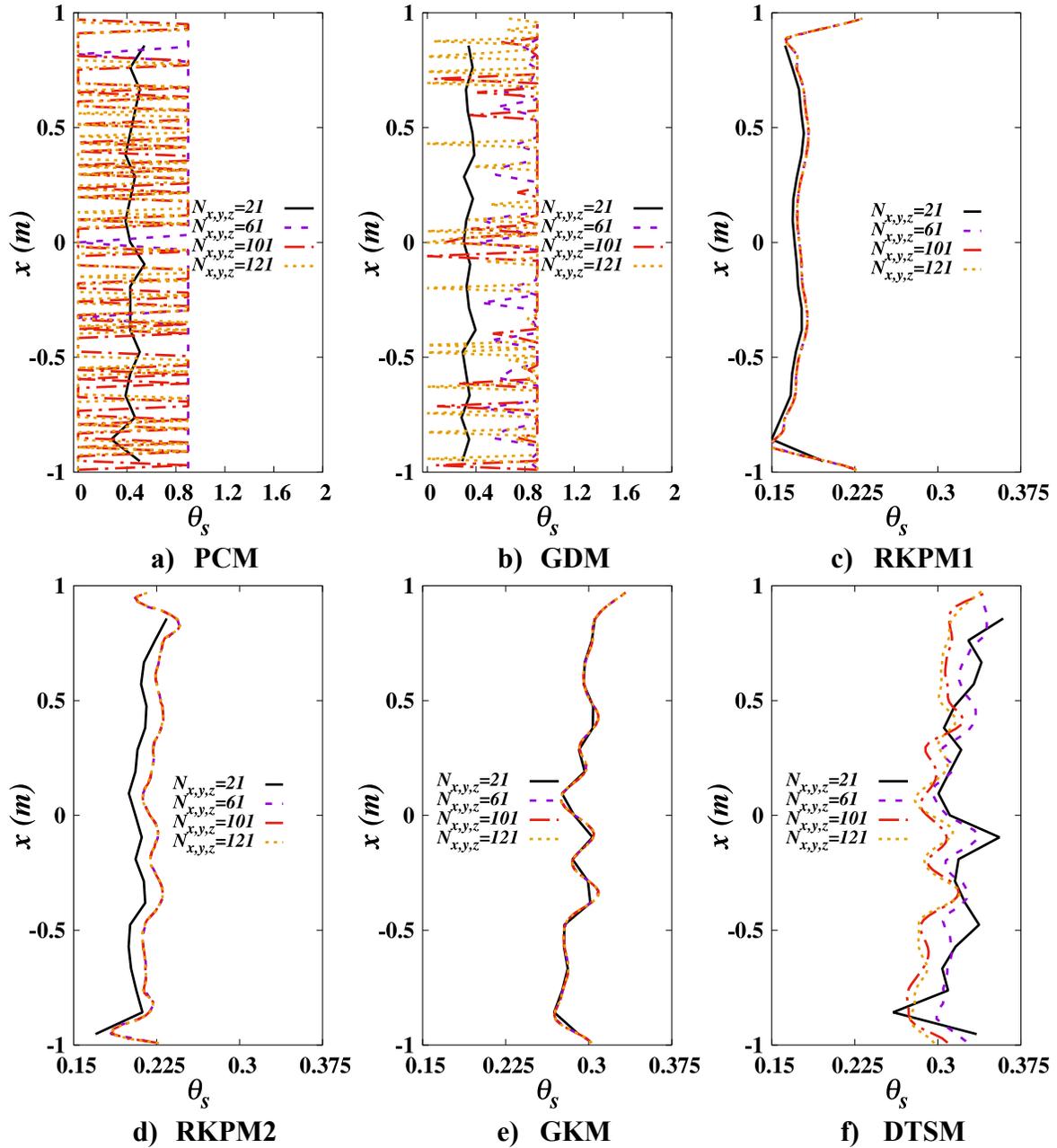

**Figure 13** Benchmark 2: The solid volume fraction along the sampling line (see figure 12) at $t = 0.0012\ s$ for different grid resolutions and CG algorithms.

The results of DTSM in figure 13f show more sensitivity to grid resolution compared to RKPMs and GKM. The level of DTSM grid-sensitivity is greater than the one observed in the first benchmark. However, for very fine grids, the results get close to each other at a slow pace since the solution of the diffusion equation gets more and more independent from its initial condition when the initial condition tends to the superposition of delta functions. This again suggests that reaching a grid-independent solution with DTSM requires a finer grid or larger $d_p/\Delta$.

*5.2.2.Other grid independency tests*

The effect of grid non-orthogonality on CG model performance for the "random particle distribution" benchmark was also investigated with the same grids introduced in table 4 (see figure 10. The $\theta_s$ predictions by different CG algorithms are reported in figure 14. This benchmark poses more challenges to CG algorithms and a noticeable level of noise can be observed in all predictions. Significant noise can destabilize EL simulations. RKPM1 shows the least amount of noise and high degrees of grid independence. Even RKPM2 has a lower noise than GKM which shows the higher sensitivity of GKM to the grid skewness. The sensitivity of GKM to grid quality and skewness stems from the approximation of the kernel integral over a cell based on the kernel value at the cell center. While the incorporation of the dynamic kernel corrector functions in RKPMs significantly improves their results compared to GKM. Among advanced models, DTSM shows larger sensitivities to the grid non-orthogonality which is traced back to the effect of the first PCM step on the final results. The cell volume and skewness in the unstructured grids considerably change locally compared to the structured grid which results in much noisier PCM predictions in the first step of DTSM on the unstructured grids. This impacts the final results of the second step as seen in figure 14d except for very fine grid resolutions.

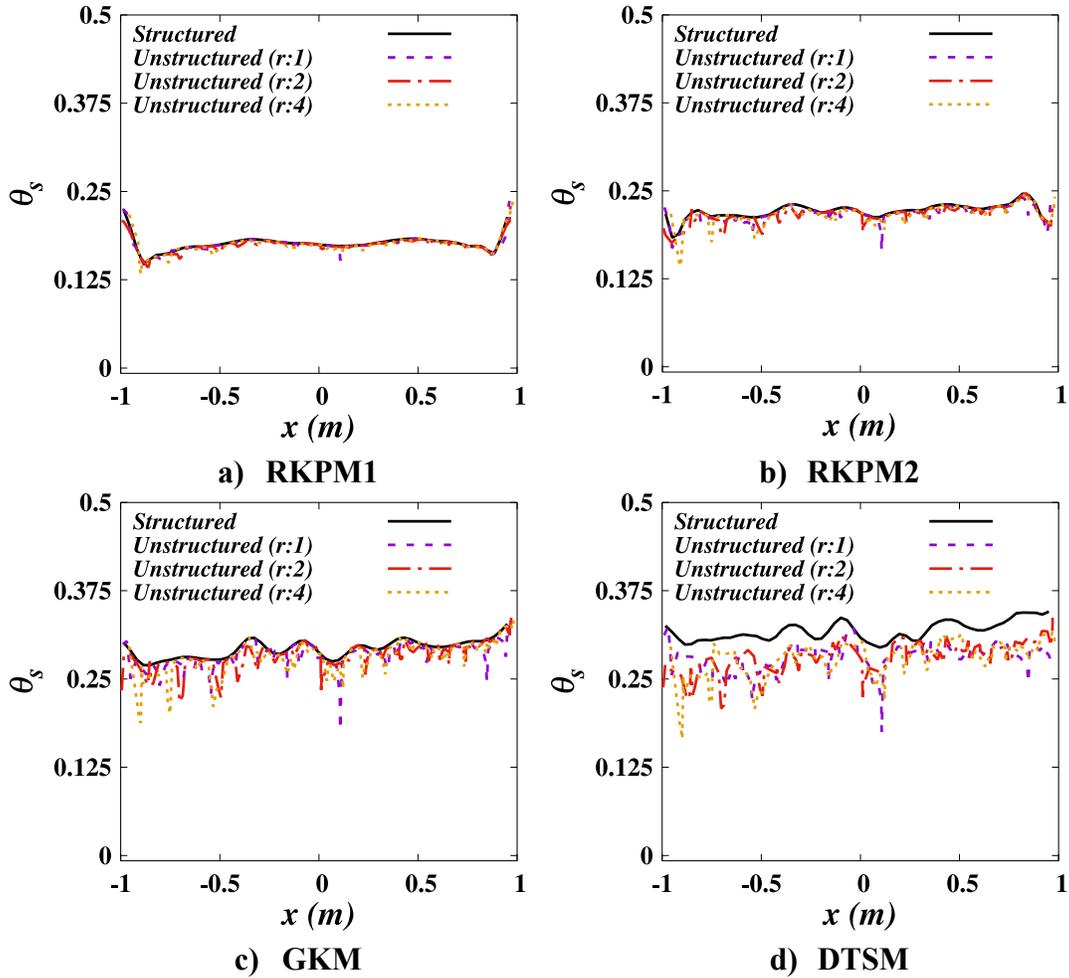

**Figure 14** Benchmark 2: The solid volume fraction along the sampling line (see figure 12) at $t = 0.0012\ s$ for different types of grid stretching and CG algorithms.

The grid non-uniformity and stretching test performed for the first benchmark in section 5.1.2 was also carried out for the "random particle distribution" benchmark and the main conclusions were similar to the first benchmark. Thus, the details of the test are not reported here for the sake of brevity; however, its quantitative performance metrics are summarized in section 5.4.

*5.3. An extended DTSM model for highly polydisperse flows*

To be able to compare the performance of different CG algorithms, they should be valid for practical EL simulations. One of the main features is the applicability to polydisperse flows. While GKM and RKPMs intrinsically account for particle polydispersity through the definition of ID diameter as a factor of particle diameter, no established and validated approach has been reported in the literature for DTSM to take the polydispersity into consideration. In the present study, we assess 3 extended DTSM variants, named "group", "constant", and "local", described in section 3.7 for this goal. In the present section, we aim to test the performance of each model against the "reference" model which solves a separate diffusion equation for each particle which is computationally prohibitive for practical problems (see section 3.7). For this purpose, we devised benchmark 3 which is similar to benchmark 2, i.e., the "random particle distribution". The differences with benchmark 2 are as follows. The number of particles is 1000 and the particle overlap is not checked for the sake of simplicity. The particle diameters satisfy a truncated Rosin-Rammler distribution with distribution parameters ($d = 0.15\ m$ and $n = 0.75$), and the minimum and maximum diameters of $0.02\ m$ and $0.2\ m$, respectively. For the "group" DTSM, three choices of group number are evaluated as $N_G = 5, 9$, and $18$. In figure 15, the number frequency of particles with $N_G = 9$ is plotted. To minimize the effect of model uncertainties and sensitivity to the first PCM step, the finest grid (grid F of table 3) with the largest $d_p/\Delta$ is chosen for this test.

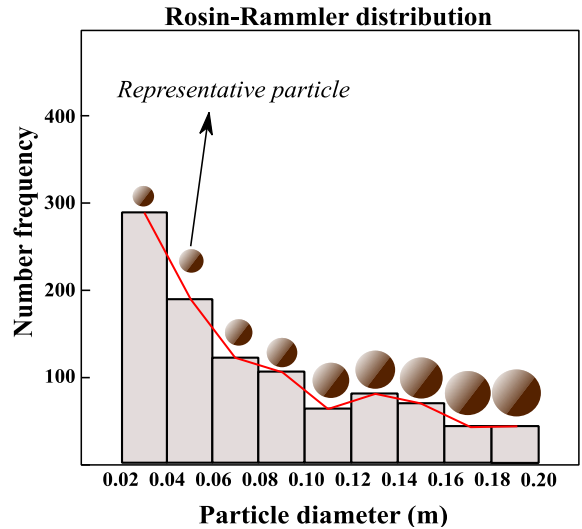

**Figure 15** Benchmark 3: The number frequency of particles with 9 bins (groups). The representative particle diameter for each group ($d_{\text{ref},g}$) is shown over each bin.

In figure 16, the results of all extended DTSM variants are compared against the reference solution at the sampling line. As it can be observed, the "constant" DTSM [23] results in a large under-prediction of $\theta_s$ along the sampling line. This is due to the overestimation of a diffusion coefficient based on the maximum particle diameter in a highly polydisperse flow. On the other hand, using the average particle diameter in the "local" DTSM approach is not a solution and leads to a large over-prediction of $\theta_s$. To analyze the reason, the scatter plot of particles and 3D solid fraction contours are illustrated in figure 17. The cause of the over-prediction of the "local" approach is envisaged considering the behavior of the CG in the z-direction where no particle exists. The absence of particles in the z-direction results in the minimum diffusion coefficient along this direction within a short distance away from $z = 0$ plane. This leads to a weak diffusion of solid fraction distribution in the z-direction for the "local" approach while the diffusion coefficient for each particle is homogeneous in the 3 spatial directions in the "reference" approach which results in considerably larger diffusion in the z-direction as seen in figure 17a. Owing to the conservativity of the algorithm, the $\theta_s$ level is much lower for the reference solution at $z = 0$ plane and the sampling line on this plane. According to figure 16, the proposed "group" DTSM can offer an acceptable compromise between the accuracy and computational cost for highly polydisperse, highly non-homogeneous flows. The error, based on Eq. (38), of cases with $N_G = 5$, 9, and 18 are 6.3%, 1.7%, and 0.7%, respectively, while the "constant" and "local" approaches have 31% and 78% errors. Therefore, for a highly-polydisperse flow, the choice of $N_G = 9$ would result in acceptable accuracy.

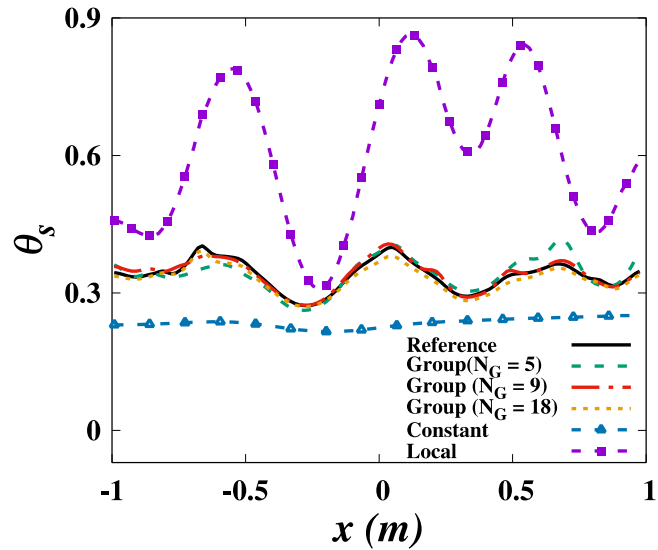

**Figure 16** Benchmark 3: The solid volume fraction along the sampling line (see figure 12) at $t = 0.0012\ s$ for the reference and extended DTSM variants.

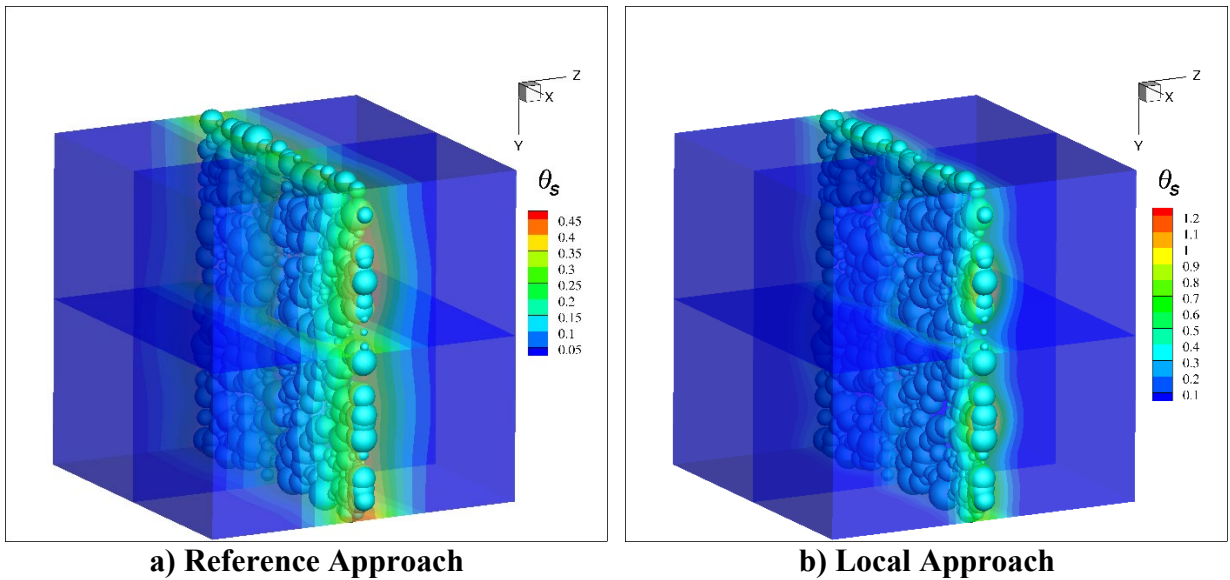

a) Reference Approach　　　　　　　　b) Local Approach

**Figure 17** Benchmark 3: The particle scatter and 3D solid volume fraction contours at $t = 0.0012\ s$ for the reference and local DTSMs.

*5.4. Quantitative performance measures*

In this section, the performance of CG algorithms for benchmarks 1-3 is summarized based on quantitative grid-independence metrics and computational cost.

figure 18 compares the grid-independence measure (error norm) defined by Eq. (38) for different CG models in the grid resolution test for benchmarks 1 and 2. The errors are computed with respect to the finest grid (grid F in table 3) on the sampling line (see figure 4). PCM measure is of the order of 1 (100% error) with an oscillating trend (even at fine grid resolutions) as the grid resolution increases. This shows the high level of grid-dependence for this CG algorithm. GDM has smaller oscillations, however, the level of error is unacceptably high and of the order of 100%. In spite of the much lower error levels of DTSM, this algorithm shows a degree of grid dependence due to the effect of the 1$^{st}$ step of the algorithm on the final solution of the diffusion equation in the 2$^{nd}$ step. The DTSM model error is a little larger, with a slower pace of reduction (smaller slope), compared to the ones of GKM and RKPMs as the grid resolution increases. RKPM1 has the lowest error (below 1% for $N/N_F > 0.1$) among all CG algorithms. RKPM1, 2, and GKM algorithms have similar performance in terms of the $C_{m \to n}^f$ measure and show the same slope of error reduction as the grid resolution increases.

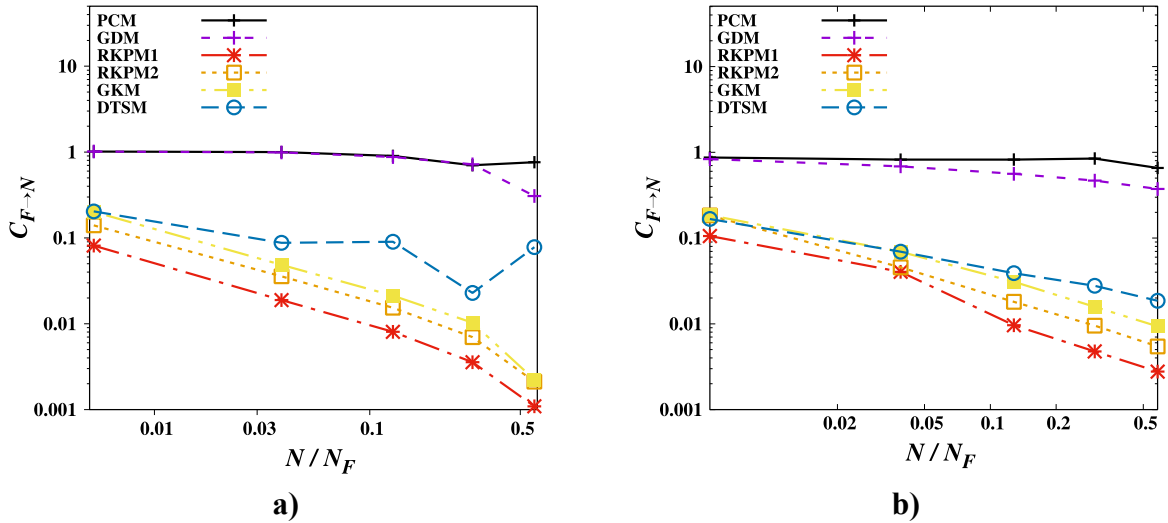

**Figure 18** The grid-independence measure versus the normalized grid resolution for different CG models in the grid resolution test for a) benchmark 1 and b) benchmark 2. $N/N_F$ is the ratio of the total number of cells in a specific grid to the one of the reference grid (grid F in table 3).

figure 19 summarizes the grid-independence measure in the grid non-orthogonality and stretching tests for benchmarks 1 and 2. Because of the high level of errors of PCM and GDM, their measures have not been reported in this figure for a better presentation. According to figure 18 and figure 19, it can be inferred that based on the grid independency property, RKPM1 is the best model followed by RKPM2 and GKM. DTSM has moderate grid-independency properties and PCM and GDM are deemed completely grid-dependent.

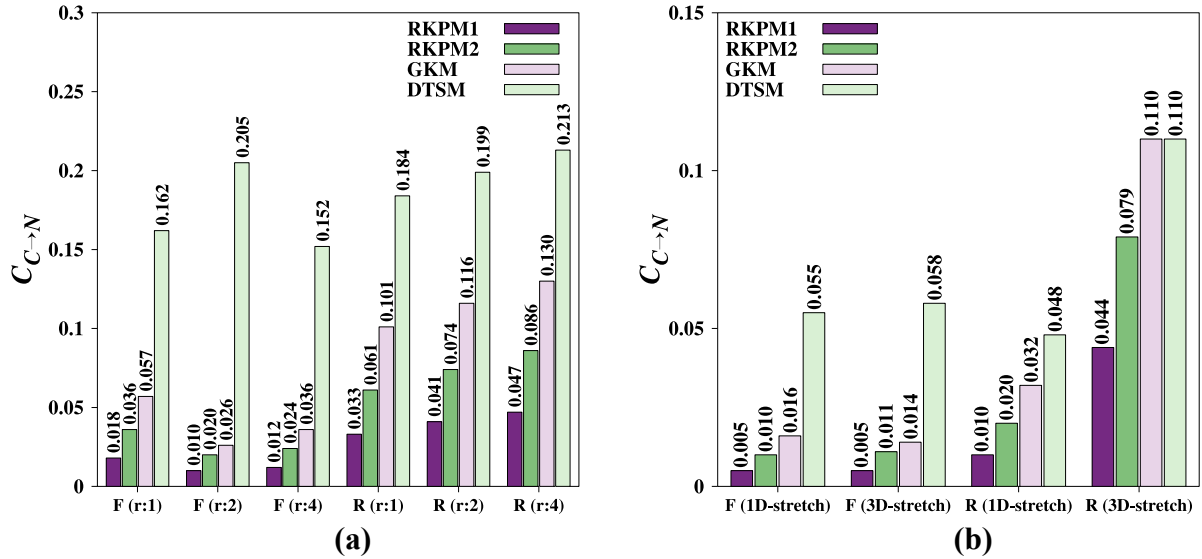

**Figure 19** The grid-independence measure for grid non-orthogonality (a) and stretching (b) tests. "F": the "falling particle", and "R": "random particle distribution" benchmarks.

The other factor determining the performance of a CG algorithm is its computational cost. The computational costs of the solution initialization and computation per Eulerian time step using different CG algorithms in benchmark 2 (monodisperse case) and benchmark 3 (polydisperse case) are reported in table 5. In addition, the performance of the efficient search algorithm to find all cells in the ID of each particle (Algorithm 8) is assessed for all models which use the ID concept, comparing the CPU time of CG models using Algorithm 8 (default) and the standard full-search algorithm at each iteration (indicated by "*" in table 5). For the monodisperse case (benchmark 2), the computational cost of each Eulerian time step dramatically reduces by the efficient search algorithm. On average, performing each time step is 22% cheaper than the cases with the standard search algorithm, and the increased overhead of the algorithm initialization is compensated for within only 4 time steps. For the highly polydisperse case (benchmark 3), the search algorithm is less efficient. For instance, it decreases the time step computation time by only 7% in the case of RKPM2. On the other hand, the initialization time rises from 1 to 284 s which requires at least 405 time steps to be compensated for by the time saved during the solution. The reason for the drop in the efficiency of the algorithm for the highly polydisperse case is the initial detection of cells in particles' ID based on the largest particle diameter which increases the initialization cost for small particles. It should be noted that Algorithm 8 is still more efficient than the standard method for all practical EL problems involving many time steps.

Table 5 The computational cost in terms of CPU time in the monodisperse (benchmark 2, 5000 particles) and polydisperse (benchmark 3, 1000 particles) "random particle distribution" on grid B ($N_{x,y,z}$ = 41). Models indicated by "*" do not use the efficient ID detection algorithm (Algorithm 8) and find all cells in each particle ID at each time step.

| Model | Benchmark 2 | | Benchmark 3 | |
|---|---|---|---|---|
| | Initialization (s) | per Eulerian time step (s) | Initialization (s) | per Eulerian time step (s) |
| PCM | <1 | 3.6 | <1 | 1.5 |
| GDM | <1 | 7.1 | <1 | 2.4 |
| RKPM1* | <1 | 25.6 | <1 | 8.1 |
| RKPM1 | 20.7 | 19.7 | 284 | 8.0 |
| RKPM2* | <1 | 27.8 | <1 | 9.7 |
| RKPM2 | 20.7 | 22.0 | 284 | 9.0 |
| GKM* | <1 | 26.8 | <1 | 9.3 |
| GKM | 20.7 | 20.8 | 284 | 8.1 |
| DTSM (constant) | <1 | 4.2 | <1 | 1.8 |
| DTSM (reference) | - | - | <1 | 466 |
| DTSM (local) | - | - | <1 | 1.9 |
| DTSM ($N_G$ = 5) | - | - | <1 | 2.8 |
| DTSM ($N_G$ = 9) | - | - | <1 | 4.6 |
| DTSM ($N_G$ = 18) | - | - | <1 | 9.5 |
| DTSM ($N_G$ = 27) | - | - | <1 | 18.0 |

Comparing the cost of EL simulation with different CG algorithms in table 5 reveals that for the monodisperse case, the computational cost per time step is 1.2, 2.0, 5.5, 5.8, and 6.1 times the one of the simple PCM for DTSM, GDM, RKPM1, GKM, and RKPM2, respectively. This advocates the current interest in DTSM for cost-demanding dense Discrete Element Method (DEM) simulations, despite its inferior grid-independence property compared to GKM or RKPMs. For the highly polydisperse case, the computational overhead per time step of all CG algorithms is close to the one in the monodisperse case, except for the standard DTSM which experiences a dramatic escalation and reaches 310 times the one of PCM. According to the data in table 5, the present "group" DTSM variant reduces the computational cost by two orders of magnitude to about 3.1 times the cost of PCM which makes DTSM still a candidate for highly polydisperse flows.

*5.5. The assessment of the novel RKPTM*

Based on the analyses performed in the preceding sections, the new RKPM1 CG model has excellent grid-independency, consistency, robustness, and smoothness. This model also has the potential to be extended to include the effect of the two-way coupling torque, according to section 3.6. This model was called RKPTM1. In this section, we aim to manifest the ability of the novel RKPTM1 model to account for the two-way coupling torque effect on the continuous-phase fluid flow. For this purpose, benchmark 4 ("falling rotating particle") is devised and introduced here. In this benchmark, a single particle of diameter $d_p = 0.125\ m$, initial velocity of $\boldsymbol{u}_p^0 = -u_{p,z}^0 \boldsymbol{k}$ ($u_{p,z}^0 = 1\ m/s$), and initial angular velocity of $\boldsymbol{\omega}_p^0 = +\omega_{p,x}^0 \boldsymbol{i}$ is released in stagnant air under the effect of gravity in the negative z-direction, drag force, and drag torque (see section 3.6 for the governing equations). The cubic computational domain and boundary conditions are the same as the ones of benchmark 1 (section 5.1). The initial position of the particle is $(0, 0, 1.15\ m)$. The computational grid is grid C of table 3. The simulations are conducted for different (non-dimensional) initial particle rotational velocities, $\omega_{p,x}^0 d_p / u_{p,z}^0$, ranging from 0 to 400 by adjusting different $\omega_{p,x}^0$ values.

The results of RKPTM1 and RKPM1 for the falling rotating particle with $\omega_{p,x}^0 d_p / u_{p,z}^0 = 400$ at an instance of simulation ($t = 0.6\ s$) are compared in figure 20. For better visualization, the corresponding animation is provided in "video2" in the supplementary materials. Without the effect of the particle rotation and drag torque on the fluid, as can be seen in the RKPM1 prediction, the pattern of the vortex formed behind the particle during the fall is axially symmetric around a vertical axis line passing through the center of the particle. The cross-section of this vortex, in a stationary frame, is shown in figure 20 (RKPM1). With the inclusion of the drag torque via RKPTM1, the induced airflow gets more complicated and asymmetric with a larger velocity magnitude on the left-hand side of the particle in the figure, which is consistent with the physics of the problem due to the rotation of the particle in the counterclockwise direction, i.e., the positive x-direction. For a better quantitative representation, the air velocity and vorticity profiles predicted by RKPTM1 along a horizontal line in the y-direction $0.5\ m$ above the particle center at $t = 0.6\ s$ are demonstrated in figure 21 for different initial particle rotational velocities. Note that the predictions of RKPM1 for all $\omega_{p,x}^0 d_p/u_{p,z}^0$ values do not differ and are the same as the RKPTM1 prediction with $\omega_{p,x}^0 d_p/u_{p,z}^0 = 0$ shown in the figure. As can be observed, at $\omega_{p,x}^0 d_p/u_{p,z}^0 = 0$, the induced air velocity profile is symmetric and in the direction of particle translation (negative z-direction). By increasing $\omega_{p,x}^0 d_p/u_{p,z}^0$, the profile gets more and more asymmetric and the velocity on the left- and right-hand sides of the particle change in accordance with the additional motion of the particle surface due to the particle rotation. This motion even induced an upward velocity in the air on the right-hand side at $\omega_{p,x}^0 d_p/u_{p,z}^0 = 400$. This is consistent with the physics due to the effect of no-slip condition on the surface of the rotating particle.

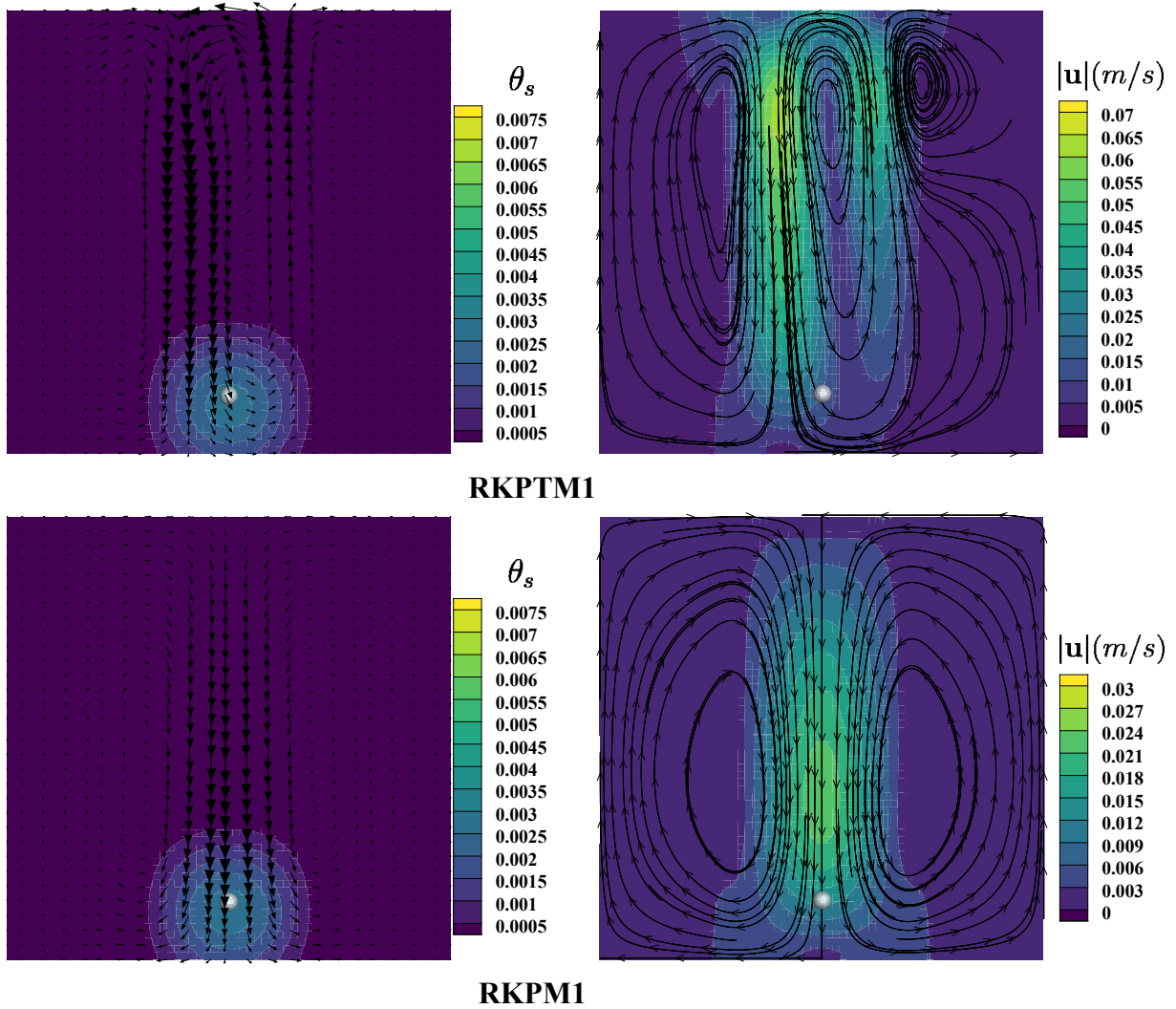

**Figure 20** The streamlines superimposed on the air velocity magnitude contours (left) and velocity vectors superimposed on the solid volume fraction contours at $t = 0.6\ s$ on the y-z plane for the case with $\omega_{p,x}^0 d_p / u_{p,z}^0 = 400$.

To further check the validity of RKPTM1 for considering the additional physics of the particle drag torque, the distribution of the two-way coupling force (per unit volume) in the surrounding air within the ID of the particle at an instance of simulation is illustrated in figure 22 for the case with $\omega_{p,x}^0 d_p / u_{p,z}^0 = 400$. The figure compares the results of RKPTM1 and RKPM1. As seen, by the addition of drag torque, the peak of force distribution shifts towards the negative y-direction to impose a net torque in the positive x-direction about the particle center which is consistent with the physics of the problem.

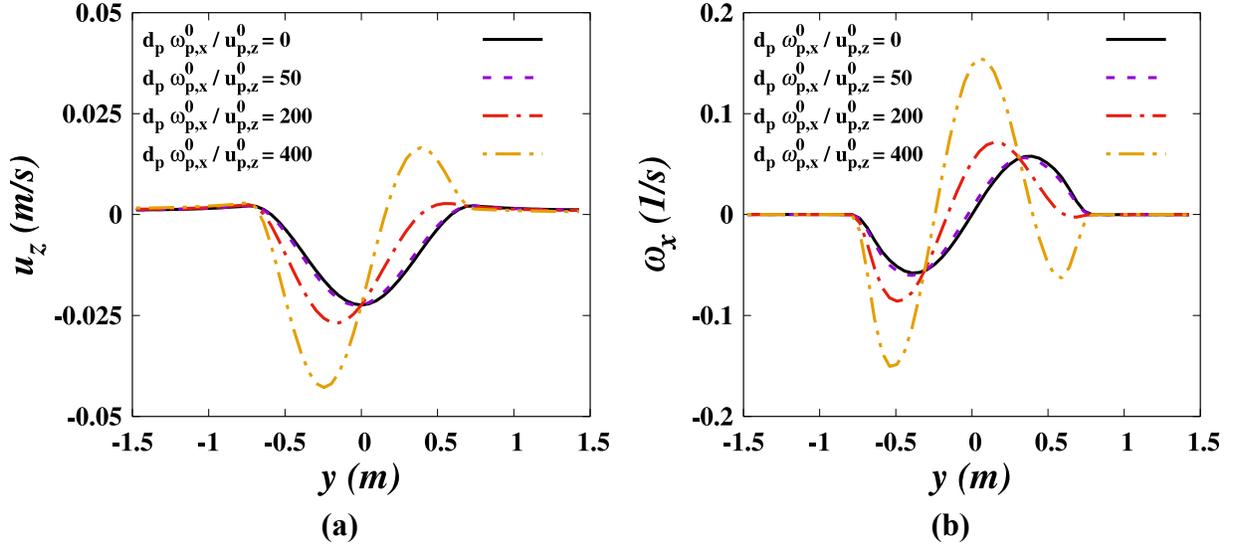

**Figure 21** Benchmark 4: The z-component velocity (a) and x-component vorticity (b) of the air flow at $t = 0.6 \; s$ along a horizontal line in the y-direction $0.5 \; m$ above the particle center. The predictions using RKPTM1 model for different (non-dimensional) initial particle rotational velocities, $\omega_{p,x}^0 d_p / u_{p,z}^0$.

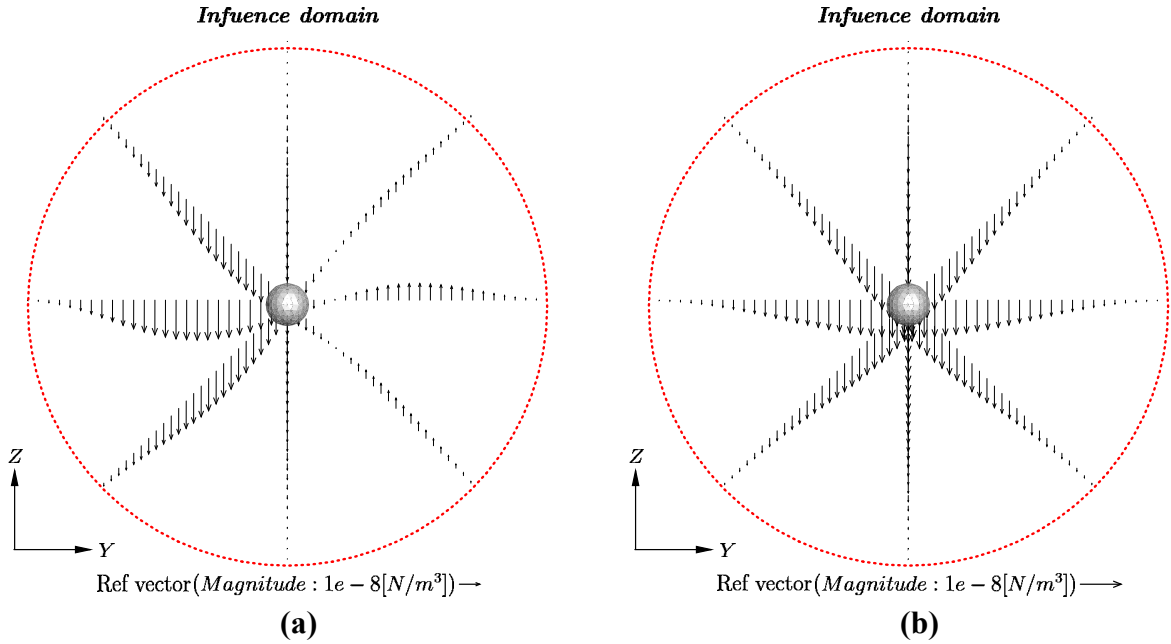

**Figure 22** Benchmark 4: The two-way coupling fluid-particle interaction force (per unit fluid mass) vector, $f_{FP}$, distribution on the y-z plane within the ID of particle at $t = 0.416 \; s$ for the case with $\omega_{p,x}^0 d_p / u_{p,z}^0 = 400$. a) RKPTM1, and b) RKPM1 (or RKPTM1 with $\omega_{p,x}^0 d_p / u_{p,z}^0 = 0$).

## 6. Conclusion

In the current work, after presenting the detailed formulation and characteristics of the widely-used CG algorithms, a new category of algorithms, called DKBM, was proposed to achieve noticeable improvements in the numerical properties (RKPM1 and 2) of CG algorithms as well as the inclusion of additional physical effects, i.e., the drag torque in RKPTM, for the two-way coupled EL simulation of particle-laden flows. An extended node-based CG model (GDM) for unstructured grids and a DTSM variant for highly polydisperse flows (G-DTSM) were also developed to be able to compare different categories of CG models under a wide range of conditions in 4 benchmarks. Several grid-independence tests, including the grid resolution, stretching, and non-orthogonality tests, were considered to assess the performance of CG models. The main characteristics of different CG algorithms based on the results of the present analyses are summarized in table 6.

Table 6 The summary of the characteristics of CG models.

| Characteristics | PCM | GDM | RKPM1 RKPTM1 | RKPM2 | GKM | DTSM G-DTSM |
|---|---|---|---|---|---|---|
| **Conservativity** | No[1] | No[1] | Yes | Yes | Yes[2] | Yes |
| **Grid-independency** | No | No | Highest | High | High | Moderate |
| **Boundness** | Yes[3] | Yes[3] | Yes | Yes | Yes | Yes |
| **Bandwidth adjustability** | No | No | Yes | Yes | Yes | Yes |
| **High-order consistency** | No | No | 1st moments, torque effect | 1st and 2nd moments | No | No |
| **Smoothness** | No | No | High | Moderate | Moderate | Moderate |
| **Robustness** | High[1] | High[1] | High[4] | Moderate[4] | High | High |
| **Boundary treatment** | Simple[5] | Simple[5] | Simple[5] | Simple[5] | Complex | Simple[5] |
| **Computational cost** | Lowest | Low | High | High | High | Moderate[6] |

1 Due to the *ad hoc* boundness enforcement
2 Needs the renormalization step
3 Needs *ad hoc* boundness enforcements
4 By hybridization with PCM
5 No specific treatment is required
6 Considering polydisperse flows (low cost for monodisperse flows)

While GDM showed improved performance compared to PCM, none of them is recommended for strongly two-way coupled EL simulations due to their major conservativity, grid-independence, and smoothness issues. On the other hand, the other 4 CG models are promising. RKPM variants had the lowest sensitivity to grid quality, skewness, and resolution, owing to the incorporation of dynamic corrector functions into their kernels. However, RKPM2 showed a more oscillating behavior and less smoothness than RKPM1 due to the nonlinearity of the corrector function in the former model. RKPMs do not need complex boundary treatments in contrast to GKM and are the only models possessing the higher-order conservativity property. The novel (hybrid) RKPM1 (and its extension RKPTM1 to account for the torque interaction) demonstrated the best performance. The hybridization of RKPMs with PCM increased the robustness of the algorithm with minor effects on its accuracy since the switching to PCM is made when $d_p/\Delta \ll 1$. In this paper, by the use of an efficient search algorithm, the cost of RKPM and GKM models reduced considerably; however, it was still higher than DTSM. The present extended DTSM, i.e., G-DTSM, is an affordable alternative to RKPM1 for computationally intensive large-scale DEM simulations. Nevertheless, it should be noted that DTSM showed higher levels of sensitivity to grid skewness and needed much finer grid resolution to offer grid-independent solutions. This was traced back to the influence of the first PCM step of the algorithm on the final coarse-grained fields. To sum up, RKPM1 and RKPTM1 are recommended for EL simulations when a higher CG operation accuracy is desirable and G-DTSM is recommended when the computational cost is the main concern, e.g., in large-scale DEM simulations.

## Appendix A

**Algorithm 1: PCM**

Initialization: set $\phi_j$ to zero
1: **for each** $p \in particlesList$ **do**
2:     $j \leftarrow$ host cell of particle $p$
3:     $\phi_{pj} = \Phi_p/V_j$
4:     **Collect:** $\phi_j = \phi_j + \phi_{pj}$
5: **end for**

### Algorithm 2: GDM

Initialization: set $\phi_j$ to zero

1: **for each** $p \in particlesList$ **do**
2:     $h_p \leftarrow$ host cell of particle $p$
3:     Decompose $h_p$ into tetrahedrons
4:     $tet_p \leftarrow$ host tetrahedron of particle $p$
5:     Compute $\omega_1, \omega_2, \omega_3, \omega_c$ coefficients based on [4]     ▷ Using Barycentric coordinate
6:     Calculate $\Phi_p^n$ and $\Phi_p^c$ by Eq. (10)
7:     **for each** $n \in$ vertices of $tet_p$ **do**
8:         $N_{pnj} \leftarrow$ neighbors of cell $h_p$ sharing vertex $n$
9:         **for each** j $\in N_{pnj}$ **do**
10:             Compute $a_{p,j}^n$ by Eq. (12)
11:             Calculate $\phi_{pj}$ by Eq. (11)
12:             **Collect:** $\phi_j = \phi_j + \phi_{pj}$
13:         **end for**
14:     **end for**
15: **end for**

### Algorithm 3: GKM

Initialization: set $\phi_j$ to zero

1: **for each** $p \in particlesList$ **do**
2:     $x_p \leftarrow$ center of particle $p$
3:     $N_c \leftarrow$ Find cells inside particle $p$ influence domain
4:     **Initialize:** $a_p = 0$
5:     **for each** $j \in N_c$ **do**
6:         $x_j \leftarrow$ center of cell $j$
7:         $V_j \leftarrow$ volume of cell $j$
8:         $VPList \leftarrow$ virtual particles list (call Algorithm 7)     ▷ Center of virtual particles
9:         Compute $g(x_j - x_p)$ by Eq. (15)
10:         **Collect:** $a_p = a_p + g(x_j - x_p)V_j$, the denominator of Eq. (14)     ▷ Collect normalization factor
11:         **if** Boundary-treatment **then**
12:             **for each** $x_{vp} \in VPList$ **do**
13:                 Compute $g(x_j - x_{vp})$ by Eq. (15)
14:                 **Collect:** $a_p = a_p + g(x_j - x_{vp})V_j$     ▷ Collect normalization factor
15:             **end for**
16:         **end if**
17:     **end for**
18:     $a_p = 1/a_p$
19:     **for each** $j \in N_c$ **do**
20:         Compute $g(x_j - x_p)$ by Eq. (15)
21:         Compute $\phi_{pj}$ by Eq. (13)
22:         **Collect:** $\phi_j = \phi_j + \phi_{pj}$     ▷ Add particle contribution
23:         **if** Boundary-treatment **then**     ▷ Check near-wall treatment
24:             **for each** $x_{vp} \in VPList$ **do**
25:                 Compute $g(x_j - x_{vp})$ by Eq. (15)
26:                 Compute $\phi_{vp,j}$ by Eq. (13)
27:                 **Collect:** $\phi_j = \phi_j + \phi_{vp,j}$     ▷ Add virtual particles contribution
28:             **end for**
29:         **end if**
30:     **end for**
31: **end for**

## Algorithm 4: RKPM1

**Initialization:** set $\phi_j$ to zero

1: **for each** $p \in particlesList$ **do**
2:     $x_p \leftarrow$ Center of particle $p$
3:     $N_c \leftarrow$ Find cells inside particle $p$ influence domain
4:     **for each** $j \in N_c$ **do**
5:        $x_j \leftarrow$ center of cell $j$
6:        **Collect:** $m_{a,b,c}^p$ by Eq. (22)
7:     **end for**
8:     Create matrix $\overline{m^p}$ by $m_{a,b,c}^p$ elements.
9:     **if** $RCond(\overline{m^p}) < 1e-3$ **then**         ▷Reciprocal condition number of m matrix
10:        Switch to PCM model         ▷ To avoid diverging
11:     **Else**
12:        $\beta_i \leftarrow$ Solve the linear Eq. (21) by Armadillo [45]
13:        **for each** $j \in N_c$ **do**
14:           $x_j \leftarrow$ center of cell $j$
15:           Compute $\phi_{pj}$ by Eqs. (16)-(20)
16:           **Collect:** $\phi_j = \phi_j + \phi_{pj}$
17:        **end for**
18:     **end if**
19: **end for**

## Algorithm 5: RKPTM

**Initialization:** set $\phi_j$ to zero

1: **for each** $p \in particlesList$ **do**
2:     $x_p \leftarrow$ Center of particle $p$
3:     $T_p \leftarrow$ Exerted torque on particle $p$
4:     Calculate $F_{p,\parallel}$ and $F_{p,\perp}$ force components by Eq. (29)
5:     $\Phi_p = F_{p,\parallel}$ is distributed over cells using Algorithm 4
6:     $x_{p*} \leftarrow$ Find modified point of action by Eq. (30)
7:     $N_c \leftarrow$ Find cells inside particle $p$ influence domain
8:     **for each** $j \in N_c$ **do**
9:        $x_j \leftarrow$ center of cell $j$
10:        **Collect:** $m_{a,b,c}^p$ and $m_{a,b,c,nx,my,qz}^p$ by Eqs. (22) and (34)
11:     **end for**
12:     Create matrix $\overline{m^p}$ by $m_{a,b,c}^p$ and $m_{a,b,c,nx,my,qz}^p$ elements based on Eq. (33)
13:     **if** $RCond(\overline{m^p}) < 1e-3$ **then**         ▷Reciprocal condition number of m matrix
14:        Switch to PCM model         ▷ To avoid diverging
15:     **Else**
16:        $\beta_i \leftarrow$ Solve the linear Eq. (33) by Armadillo [45]
17:        **for each** $j \in N_c$ **do**
18:           $x_j \leftarrow$ center of cell $j$
19:           Compute $\phi_{pj}$ by Eqs. (16)-(20) and $\Phi_p = F_{p,\perp}$
20:           **Collect:** $\phi_j = \phi_j + \phi_{pj}$
21:        **end for**
22:     **end if**
23: **end for**

## Algorithm 6: DTSM

Initialization: set variables $\phi_j$ and $\phi_{j,g}$ to zero

1: **for** each $p \in particlesList$ **do**
2:     $g \leftarrow$ group of particle $p$                                      ▷ Particles are classified into groups
3:     $j \leftarrow$ host cell of particle $p$
4:     **Collect:** $\phi_{j,g} = \phi_{j,g} + \Phi_p/V_j$                          ▷ PCM model (the first step)
5: **end for**
6: $particleGroupsList \leftarrow$ Define particle diameter classes based on the $D_f$ estimating approach
7: **for** each $g \in particleGroupsList$ **do**
8:     $D_f \leftarrow$ Diffusion coefficient for group $g$ based on table 1
9:     Solve Eq. (35) for $\phi_{j,g}$ with pre-stored initial condition from the first step
10:     **Collect:** $\phi_j = \phi_j + \phi_{j,g}$
11: **end for**

## Algorithm 7: Near wall treatment for GKM model

Inputs: particle position ($x_p$), cell $c$ coordinate ($x_c$), and mesh information including walls and cells coordinates
Outputs: Virtual particles positions (*VPList*)

1: $(L_{c,w}, N_{c,w})$ = Call Sub-Algorithm 1                        ▷Triggered only at solution startup
2: **Repeat**
3:     $x_p \leftarrow$ particle $p$ coordinate
4:     $x_c \leftarrow$ cell $c$ coordinate
5:     **for** each $w \in wallsList$ **do**
6:        $x_{vp} \leftarrow$ Call Sub-Algorithm 2($L_{c,w}, N_{c,w}, x_p, x_c$)      ▷The first layer of virtual particles
7:        **if** $x_{vp} \notin VPList$ **then**
8:           Add $x_{vp}$ into $VPList$                                   ▷Add the first layer
9:           **for** each $w' \in wallsList$ **do**
10:             **if** ($w' \neq w$) **then**
11:                $x'_{vp} \leftarrow$ Call Sub-Algorithm 2($L_{c,w}, N_{c,w}, x_{vp}, x_c$)    ▷the second layer of virtual particles
12:                **if** $x'_{vp} \notin VPList$ **then**
13:                    Add $x'_{vp}$ into $VPList$                         ▷Add the second layer
14:                    $x_p \leftarrow x'_{vp}$ then go to line 3
15:                **end if**
16:             **end if**
17:           **end for**
18:        **end if**
19:     **end for**
20: **Until**: no new virtual particle is found
21: Return $VPList$

## Sub-Algorithm 1 (Triggered: solution startup)

Inputs: Mesh information including walls and cells coordinates
Outputs: The cell to wall distance vector ($L_{c,w}$) and wall normal vector ($N_{c,w}$)

1: **for** each $c \in cellsList$ **do**
2:     $x_c \leftarrow$ cell $c$ coordinate
2:     **for** each $w \in wallsList$ **do**
3:        **Set:** $L_{c,w} \leftarrow$ Loop over all faces ($f$) of $w$ and find the minimum $|x_c - x_f|$
4:        **Set:** $N_{c,w} \leftarrow$ The face normal unit vector of the face used to calculate $L_{c,w}$
5:     **end for**
6: **end for**

## Sub-Algorithm 2 (Triggered: Run-time)

Inputs: Cell to wall vector distance ($L_{c,w}$), walls normal vector ($N_{c,w}$), real particle position ($x_p$), and cell coordinate ($x_c$)
Outputs: Virtual particle coordinate $x_{vp}$

1:    $\delta_p \leftarrow$ Influence domain size
2:    **if** $|L_{c,w}| < \delta_p/2$ **then**
3:        **Set:** $L_{p,c} = x_p - x_c$      ▷ Particle $p$ to cell distance vector
4:        **Set:** $L_{p,w} = L_{p,c} + L_{c,w}$      ▷ Particle-wall distance vector
5:        **Set:** $d_{p,w} = N_{p,c}.L_{p,w}$      ▷ Particle-wall normal distance
6:        **if** ($d_{p,w} < \delta_p/2$) **then**
7:           Return $x_{vp} = x_p + 2L_{p,w}$      ▷ Return location of virtual particle
8:        **end if**
9:    **end if**

## Algorithm 8: Cell Search Algorithm (Triggered: Startup)

1: $d_{p,\max} \leftarrow$ maximum particle diameter
2: $R_b \leftarrow$ ratio of $\delta$ to particle diameter
3: $\delta_{\max} = d_{p,\max} R_b$      ▷ Influence domain global size
4: **for each** $c \in cellsList$ **do**
5:     $N_c \leftarrow$ List of neighbor cells of cell $c$
6:     **Set:** $\Delta_c \leftarrow V_c^{1/3}$      ▷ Cell length scale
7:     **Set:** $\delta_c = \delta_{\max} + \Delta_c$      ▷ Local influence domain size
8:     $verticesList \leftarrow$ cell $c$ vertices
9:     **for each** $v \in verticesList$ **do**
10:       $N_c^v \leftarrow$ Find cells around $v$ considering influence domain size $\delta_c$      ▷ Find vertex $v$ neighbors
11:       **Collect:** $N_c = N_c + N_c^v$ cell $c$ neighbors      ▷ Collect vertex $v$ neighbors
12:     **end for**
13:     $N_c \leftarrow$ unique($N_c$)      ▷ Remove duplicate cells
14: **end for**

## Algorithm 9: OpenFOAM Lagrangian step

1: $\Delta T_E \leftarrow$ Eulerian time-step
2: Particle injection
3: **for each** $p \in particlesList$ **do**
4:     **Set:** $\Delta T_L = 0$,
5:     **While** $\Delta T_L \leq \Delta T_E$ **do**
6:       **Set:** $d_E = (\Delta T_E - \Delta T_L)|u_p|$      ▷ The maximum displacement
7:       $d_{Face} \leftarrow$ Distance observed to hit a cell face
8:       $Co_L \leftarrow$ Lagrangian Courant number
9:       **Set:** $d_{Co} = Co_L \times \Delta_{host}$      ▷ Distance travelled based on the Courant number
10:       **Set:** $\Delta t_p = \min(d_E/|u_p|, d_{Face}/|u_p|, d_{Co}/|u_p|)$      ▷ Lagrangian time-step
11:       Update particle $p$ location ($x_p$) by $\Delta t_p$ and $u_p$
12:       Interpolating fluid variable $\mu_c$, $u_c$ and $\rho_c$ into particle $p$ location
13:       Update $u_c$ using interphase force by Eq. (37)
14:       **Set:** $F_{p,\text{tot}} = F_{ncp} + F_{cp}$      ▷ Calculate coupled and non-coupled forces on the particle
15:       Update $u_p$ by integrating of forces over time
16:       Compute $\Phi_p$ ($F_p$ or $V_p$)      ▷ Determine particle two-way variables
17:       Call CG algorithm with $\Phi_p = \text{frac} \times \Phi_p$, frac $= \Delta t_p/\Delta T_E$      ▷ The first step of CG algorithm in case of DTSM
18:       **Collect:** $\Delta T_L = \Delta T_L + \Delta t_p$
19:     **end while**
20: **end for**
21: Call the second step of DTSM      ▷ Alternative location for calling all CG algorithms (with frac = 1) instead of line 17